\documentclass[prd,aps,lengthcheck,twocolumn,nofootinbib,notitlepage,floatfix,superscriptaddress]{revtex4-2}
\usepackage{graphics,graphicx}
\usepackage{amsmath, amssymb}
\usepackage{multirow}
\usepackage{color}
\usepackage{verbatim}
\usepackage{hyperref}
\usepackage[normalem]{ulem}
\usepackage{ulem}
\usepackage{color}
\usepackage{cancel}
\usepackage{mathtools}

\def\fm3{\;\text{fm}^{-3}}

\renewcommand\sout{\bgroup\color{blue} \ULdepth=-.5ex \ULset}
\def\slashchar#1{\setbox0=\hbox{$#1$}  
\dimen0=\wd0     
\setbox1=\hbox{/} \dimen1=\wd1  
\ifdim\dimen0>\dimen1   
\rlap{\hbox to \dimen0{\hfil/\hfil}} 
#1     
\else     
\rlap{\hbox to \dimen1{\hfil$#1$\hfil}} 
/      
\fi}

\newcommand{\eps}{\epsilon}

\newcommand{\pp}{\partial}

\begin{document}
\title{Suppression of dynamical momentum-space shell by chiral symmetry}

\author{Bikai Gao}
\email{bikai@rcnp.osaka-u.ac.jp}
\affiliation{Research Center for Nuclear Physics (RCNP), Osaka University, Osaka 567-0047, Japan}

\author{Micha\l{} Marczenko}
\email{michal.marczenko@uwr.edu.pl}
\affiliation{Incubator of Scientific Excellence - Centre for Simulations of Superdense Fluids,
University of Wroc\l{}aw, plac Maksa Borna 9, 50204 Wroc\l{}aw, Poland}

\date{\today}

\begin{abstract}
We investigate the appearance of quark degrees of freedom in dense isospin-symmetric nuclear matter. We employ the parity doublet model to incorporate chiral dynamics. Specifically, we contrast quarkyonic matter, in which quarks occupy states above the nucleon Fermi surface, with baryquark matter, in which quarks populate states inside the nucleonic Fermi sea. We find that while baryquark matter is generally energetically favored over quarkyonic matter, the self-consistent treatment of the momentum-space shell reveals that purely hadronic matter provides the lowest free energy up to densities well beyond nuclear saturation. Consequently, the contribution of quarks is not relevant within the model’s domain of applicability, even though chiral symmetry becomes restored. This demonstrates that the onset of quark degrees of freedom and the restoration of chiral symmetry need not coincide.
\end{abstract}

\maketitle

\section{Introduction}
Understanding the dense nuclear matter remains one of the central challenges in nuclear physics and astrophysics. Recent advances in multi-messenger astronomy, particularly the detection of neutron stars with masses exceeding two solar masses~\cite{Miller:2021qha, Riley:2021pdl} and the gravitational wave observations from neutron star mergers~\cite{LIGOScientific:2017vwq, LIGOScientific:2017ync}, have improved our understanding of ultra-dense matter. These astrophysical observations have placed unprecedented constraints on theoretical models incorporating the microscopic physics of dense nuclear matter equation of state. This has led to the development of various unified approaches that can describe both hadronic and quark degrees of freedom within a single framework. They include chiral symmetry restoration, hadronic interactions, and the possible emergence of quark degrees of freedom with associated phase transitions at higher densities. The transition region from hadronic to quark matter, occurring approximately between $2n_0$ and $5n_0$ (where $n_0$ represents the nuclear saturation density), presents significant theoretical challenges due to the non-perturbative nature of quantum chromodynamics (QCD) in this regime.

Traditional approaches have often relied on separate descriptions for hadronic and quark matter, with an abrupt phase transition between the two regimes~\cite{Lenzi:2012xz, Benic:2014jia, Gartlein:2023vif, Carlomagno:2023nrc, Ayriyan:2024zfw, Gao:2024lzu, Christian:2025dhe, Yang:2025iyv}. However, recent theoretical evidence suggests that the transition may be a more gradual process, potentially involving a crossover or mixed phase~\cite{Masuda:2012kf, Masuda:2012ed, Kojo:2014rca, Baym:2017whm, Baym:2019iky, Blaschke:2021poc, Kojo:2021wax, Minamikawa:2020jfj, Kong:2023nue, Kong:2025dwl}. This has led to the development of various unified approaches that can describe both hadronic and quark degrees of freedom within a single framework.

The concept of quarkyonic matter offers an intriguing framework for understanding this transition~\cite{McLerran:2007qj}. The basic concept is that at sufficiently high baryon chemical potential, the degrees of freedom inside the Fermi sea can be treated as quarks while confining forces remain important only near the Fermi surface~\cite{McLerran:2008ua, Hidaka:2008yy, Fukushima:2015bda, Duarte:2021tsx, Kojo:2021ugu, McLerran:2018hbz, Jeong:2019lhv, Sen:2020peq, Fujimoto:2023mzy, Fujimoto:2024doc, Kojo:2024ejq, Ivanytskyi:2025cnn, Gao:2025rbq, Kojo:2025vcq, Fujimoto:2025trx}. In this picture, baryons emerge as correlations between quarks at the Fermi surface. This creates a distinctive momentum-space structure where a Fermi sea of deconfined quarks is surrounded by a shell of confined baryons. Recent works in Refs.~\cite{Koch:2022act, Poberezhnyuk:2023rct, Koch:2025sfv} have explored an alternative configuration called baryquark matter. In this scenario, the momentum-space structure is inverted, and a Fermi sea filled with confined baryons is surrounded by a shell of deconfined quarks. Through energy minimization, they demonstrated that this baryquark configuration may be energetically favored over quarkyonic matter in systems with constituent quarks and hard-core repulsion. This finding raises important questions about the nature of dense QCD matter and the mechanisms governing the hadron-quark transition.

Simultaneously, models incorporating chiral symmetry restoration, such as the parity doublet model~\cite{Detar:1988kn, Jido:2001nt}, have shown significant success in describing hadronic matter at finite temperature and density~\cite{universe5080180, PhysRevC.100.025205, Mukherjee:2017jzi, Minamikawa:2020jfj, Marczenko:2021uaj, Marczenko:2022hyt, Kong:2023nue, Gao:2024chh, Gao:2024lzu, Marczenko:2020jma, Eser:2023oii, Eser:2024xil, Shao:2022njr, Ma:2023eoz, Guo:2023mhf, Guo:2023som, Ma:2023ugl, Gao:2024mew, Marczenko:2023ohi, Koch:2023oez, Marczenko:2024nge}. The parity doublet model introduces a chiral invariant mass component for baryons that persists even when chiral symmetry is restored. This feature enables a more realistic description of the nucleon mass evolution with increasing density, which is crucial for understanding the equation of state of dense matter. This approach was extended by combining the quarkyonic matter framework with the PDM, introducing a chiral invariant mass for both baryons and constituent quarks~\cite{Gao:2024jlp}. Such integration enables a consistent treatment of both the confinement properties and chiral symmetry aspects in dense nuclear matter.

In this paper, we study the equation of state of dense isospin symmetric nuclear matter by exploring both quarkyonic and baryquark matter configurations with chiral symmetry restoration effects incorporated via the parity doublet model. We investigate how the self-consistent treatment of the momentum-space shell affects the energetics of these configurations and the onset of quark degrees of freedom. Our key findings include

This paper is organized as follows: In Sec.\ref{sec_formulation}, we explain the formulation of
the present model. In Sec.~\ref{sec:delta1}, we discuss the results obtained with an ansatz for the width of the momentum-space shell. In Sec.~\ref{sec:delta2}, we treat the width of the momentum-space shell self-consistently and discuss our findings. Finally, in Sec.~\ref{sec:summary}, we summarize our results.

\section{Formulation}
\label{sec_formulation}

In this section, we briefly review the construction of the relativistic mean-field model based on the $\rm SU(2)$ parity doublet structure for the nucleon. In the parity doublet framework, the excited nucleon $N(1535)$ with negative parity is regarded as the chiral partner of the ground state nucleon $N(939)$ with positive parity. Assuming isospin symmetry, the thermodynamic potential in the mean-field approximation at zero temperature is given by~\cite{Gao:2024chh, Minamikawa:2020jfj}:
\begin{equation}\label{eq:omega_pdm}
\Omega= \Omega_F + V(\sigma) - \frac{1}{2} m_{\omega}^{2} \omega^{2} \textrm,
\end{equation}
where
\begin{equation}
\Omega_{F} = \sum_{\alpha=\pm}-4\int\limits_0^{k_F} \frac{\mathrm{d}^3 \mathbf{k}}{(2 \pi)^{3}}\left(\mu_\alpha-E_{\alpha}\right) \textrm,
\end{equation}

and the $\sigma$ mean-field potential $V(\sigma)$ is given by
\begin{equation}
V(\sigma) = -\frac{1}{2}\bar{\mu}^{2}\sigma^{2} + \frac{1}{4}\lambda_4 \sigma^4 -\frac{1}{6}\lambda_6\sigma^6 - m_{\pi}^{2} f_{\pi}\sigma \textrm, \\
\end{equation}

Here, $\alpha = \pm$ denotes the positive-parity nucleon and its negative-parity chiral partner. The pion decay constant $f_{\pi}=92.4$~MeV, and $E_\alpha = \sqrt{\mathbf{k}^2 + m_\alpha^2}$ represents the energy of nucleons with mass $m_\alpha$ and momentum $\mathbf{p}$.
The effective masses are given as
\begin{equation}\label{eq:pdm_mass}
m_{\pm} = \sqrt{m_0^2 + \left(\frac{g_1 + g_2}{2}\right)^2\sigma^2} \mp \frac{g_1 - g_2}{2}\sigma,
\end{equation}
where $g_1$ and $g_2$ are coupling constants determined by the vacuum values of $m_\pm$ and the chiral invariant mass $m_0$. As the density increases, chiral symmetry restoration drives the mean field $\sigma$ toward zero, causing the positive and negative parity nucleon masses to degenerate at $m_0$. The effective chemical potential $\mu_\pm = \mu_B - g_\omega \omega$, where $g_\omega$ is the model parameter. In the current model, the positive- and negative-parity states are coupled to the vector meson $\omega$ with the same strength, $g_\omega$. The parameters $\bar{\mu}^2$, $\lambda_4$, $\lambda_6$, and $g_\omega$ are fixed by the properties of the nuclear ground state at vanishing temperature and density $n_0= 0.16~\rm fm^{-3}$. Following the method outlined in Refs.~\cite{Motohiro, Minamikawa:2020jfj, Gao:2024chh} for different values of $m_0$. We note that a relation exists between the chiral invariant mass $m_0$ and the stiffness of the equation of state~\cite{PhysRevC.100.025205, Minamikawa:2020jfj}. Namely, for small values of $m_0$, large values of the scalar couplings are needed to account for the nucleon mass at saturation, which in turn requires larger values of the repulsive coupling $g_\omega$ due to the equilibrium state at the saturation density. As baryon density increases, the contribution of the $\omega$ mean field becomes dominant and the equation of state becomes stiffer. 

When quarks saturate the low–energy states, Pauli blocking suppresses the formation of excited baryon states, thereby pushing the onset of heavier baryonic excitations to larger chemical potentials. As a consequence, within this description the negative-parity state $N(1535)$ does not emerge when the baryon chemical potential reaches its mass, but only at higher values. Furthermore, at finite $N_c$ the validity of the quarkyonic (and baryquark) picture is restricted to a window of quark chemical potential, $\Lambda_{\text{QCD}} < \mu_q < \sqrt{N_c}\,\Lambda_{\text{QCD}}$, with the upper bound $\mu_B=3\mu_q$ lying close to the $N(1535)$ mass. For these reasons, we do not include the negative-parity state in the density region considered in this work.

For our considerations, we fix the the chirally invariant mass $m_0 = 800~$MeV. The qualitative features discussed below, however, are robust against variations of $m_0$. In this work, we restrict the current study to the density region where the negative-parity state is not populated, i.e., $n_N \lesssim 8n_0$. This allows us to neglect its contribution to the thermodynamic potential. Hereafter, we denote the positive-parity nucleon with a subscript $N$.

\begin{figure}
    \centering
    \includegraphics[width=\linewidth]{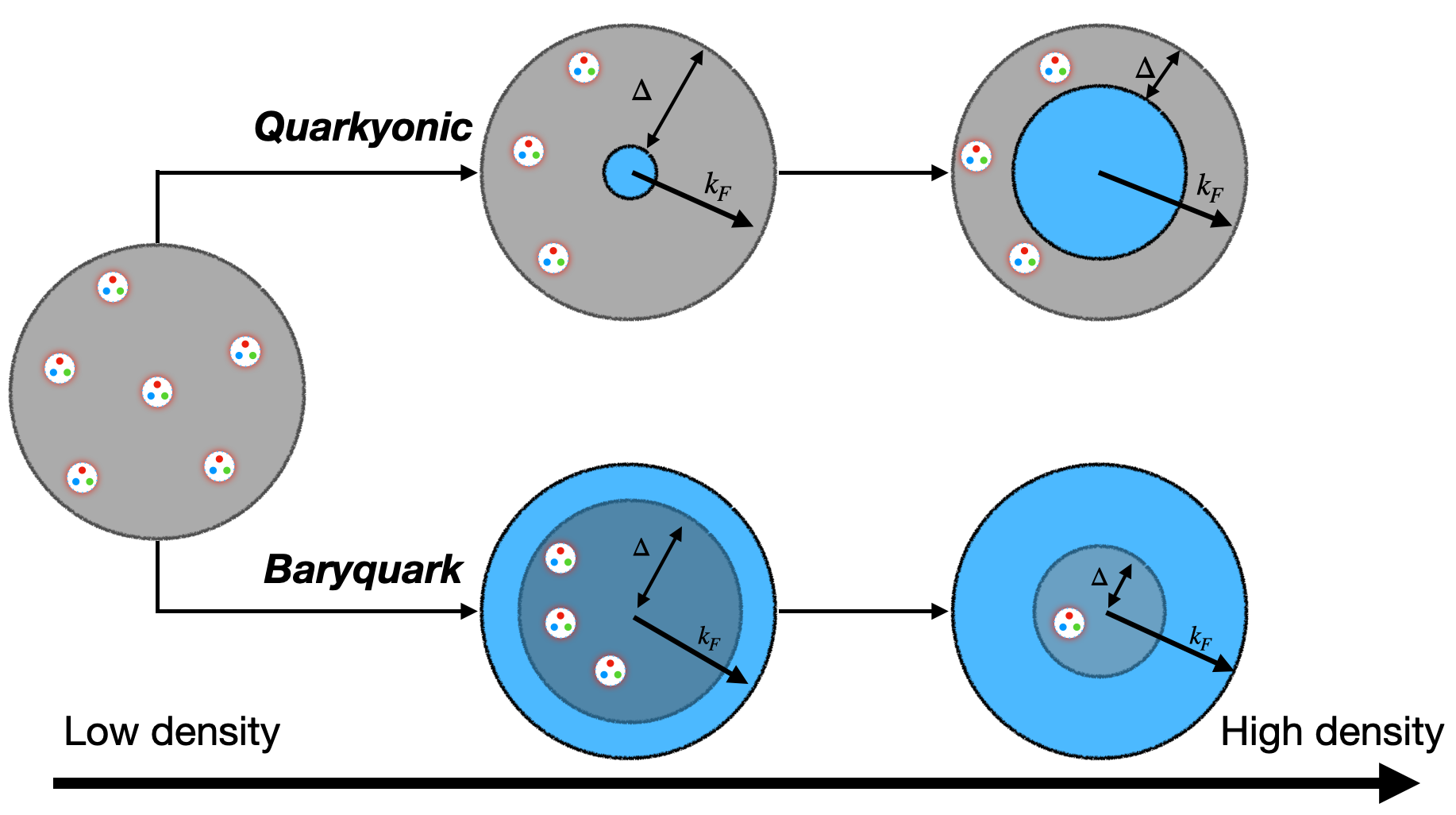}
    \caption{Schematic plot showing the structure of the Fermi sea for quarkyonic matter (upper) and the baryquark matter (lower) with increasing density. Nucleons exist in the momentum shell $\Delta$ in the outer region of the Fermi sea in the quarkyonic phase. In the baryquark picture, nucleons exist in the inner core of the Fermi sea.}
    \label{fig:mnechanism}
\end{figure}

\subsection{Quarkyonic matter}

In the quarkyonic matter scenario, nucleons occupy a momentum shell region near the Fermi surface between momenta $k_F - \Delta$ and $k_F$, where $\Delta$ represents the momentum shell width. This is depicted in Fig.~\ref{fig:mnechanism}. Consequently, quarks are surrounded by the shell of baryons and occupy momenta up to $\left(k_F - \Delta\right)/N_c$, where $N_c=3$ is the number of color degrees of freedom. The kinetic part $\Omega_F$ of the thermodynamic potential in Eq.~\eqref{eq:omega_pdm} is replaced by
\begin{equation}\label{eq_omega_qc}
\Omega_F = \Omega_N + \Omega_q
\end{equation}
where the thermodynamic potentials of the nucleons and quarks are
\begin{equation}
\Omega_N = -4 \int\limits^{k_F}_{k_F - \Delta } \frac{\mathrm{d}^3 \mathbf{k}}{(2 \pi)^{3}}\left(\mu_N-E_N\right) \textrm,
\end{equation}
and
\begin{equation}
\Omega_q = - 4N_c \int\limits_{0}^{ (k_F - \Delta)/N_c} \frac{\mathrm{d}^3 \mathbf{k}}{(2 \pi)^{3}}(\mu_q - E_q),
\end{equation}
respectively. The quark chemical potential $\mu_q = \mu_B / N_c$.

The quark single-particle energy is given by
\begin{align}
E_q = \sqrt{{\bf k}^2 + m_q^2},
\end{align}
with $m_q$ representing the constituent quark mass.
The baryon number density can be derived from the pressure through the thermodynamic relation \mbox{$n_B = -\partial \Omega / \partial \mu_B$}, yielding:
\begin{equation}\label{eq_numberdensity}
n_{B} = n_N + n_q\textrm,
\end{equation}
with
\begin{equation}
n_N = \frac{2}{3\pi^2}\left(k_F^3 - \left(k_F - \Delta \right)^3 \Theta(k_F - \Delta)\right)
\end{equation}
and
\begin{equation}
n_q = \frac{2}{3\pi^2} \left(\frac{k_F - \Delta}{N_c}\right)^3 \Theta(k_F - \Delta),
\end{equation}
where $\Theta(x)$ is the Heaviside step function.
The quarkyonic model features regimes of distinct behavior depending on the baryon density, $n_B$. At low densities, the Fermi momentum $k_F$ is smaller than the shell width $\Delta$ and the step function $\Theta(k_F - \Delta)$ vanishes. This corresponds to the purely hadronic matter with no quark degrees of freedom. As the density increases, $k_F$ eventually exceeds $\Delta$ and the system transitions to the quarkyonic phase, characterized by the emergence of a non-zero quark Fermi momentum, $(k_F - \Delta)/N_c$.

\subsection{Baryquark matter}

In the baryquark scenario, the momentum-space structure is inverted compared to quarkyonic matter (see Fig.~\ref{fig:mnechanism}). Namely, nucleons occupy the Fermi sea up to momentum $\Delta$, while quarks populate the shell region with momenta between $\Delta/N_c$ and $k_F/N_c$. In this case, the kinetic part $\Omega_F$ of the thermodynamic potential in Eq.~\eqref{eq:omega_pdm} is replaced by
\begin{equation}\label{eq_omega_bc}
\Omega_F = \Omega_N + \Omega_q
\end{equation}
where the thermodynamic potentials of the nucleons and quarks are given by
\begin{equation}
\Omega_N = -4\int\limits_0^\Delta \frac{\mathrm{d}^3 \mathbf{k}}{(2 \pi)^{3}}\left(\mu_N-E_N\right) \textrm,
\end{equation}
and
\begin{equation}
\Omega_q = - 4N_c \int\limits_{\Delta/N_c}^{k_F/N_c} \frac{\mathrm{d}^3 \mathbf{k}}{(2 \pi)^{3}}(\mu_q - E_q),
\end{equation}
respectively.

Consequently, the nucleon and quark densities are given by
\begin{equation}
n_N = \frac{2}{3\pi^2}k_F^3 \Theta(\Delta - k_F) + \frac{2}{3\pi^2}\Delta^3 \Theta(k_F - \Delta)
\end{equation}
and
\begin{equation}
n_q = \frac{2}{3\pi^2}\left(\left(\frac{k_F}{N_c}\right)^3 - \left(\frac{\Delta}{N_c}\right)^3\right) \Theta(k_F - \Delta) \textrm,
\end{equation}
respectively.

In the baryquark model, the system transitions from pure hadronic matter to the mixed phase when the Fermi momentum $k_F$ exceeds $\Delta$. Below this threshold ($k_F < \Delta$), the system reduces to pure hadronic matter with density $n_B = 2k_F^3/3\pi^2$.
The constituent quark mass is assumed to be the same in both quarkyonic and baryquark models~\cite{Gao:2024jlp}:
\begin{equation}\label{eq:quark_mass}
m_q = \frac{m_N}{3} \textrm,
\end{equation}
where $m_N$ is given through Eq.~\eqref{eq:pdm_mass} for the positive-parity nucleon. This mechanism naturally implements the duality between hadronic and quark degrees of freedom through the medium-dependent mass for nucleons and constituent quarks, as well as through the chirally invariant mass.

\begin{figure}[t!]\centering
\includegraphics[width=1\hsize]{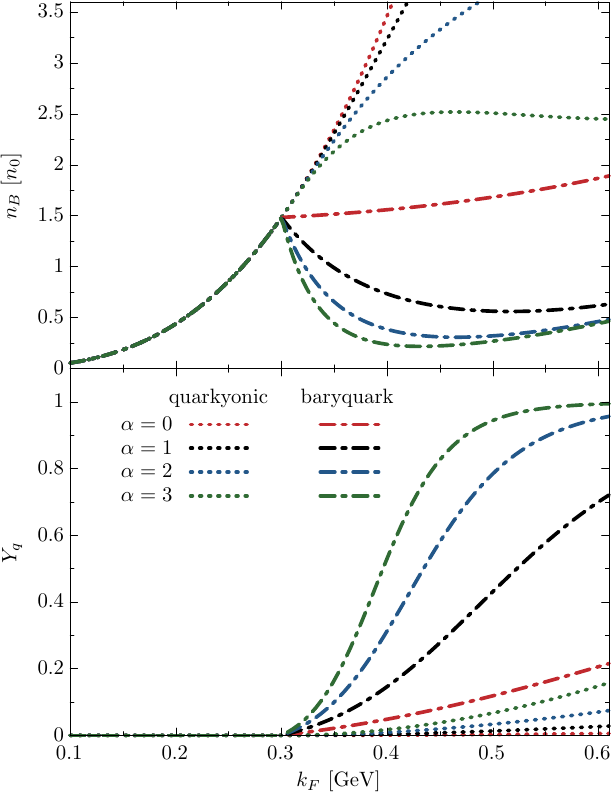}
\caption{Normalized baryon number density $n_B$ (top panel) and the quark fraction $Y_q$ (bottom panel) as functions of Fermi momentum $k_F$ for different choices of the $\alpha$ parameter.}
\label{fig_nb_kf}
\end{figure}

\section{Parametrized momentum-space shell}
\label{sec:delta1}

The momentum shell width $\Delta$  is a key parameter in both quarkyonic and baryquark descriptions of nuclear matter. Previous studies of quarkyonic matter introduced different parametrizations for $\Delta$ as a function of the Fermi momentum~\cite{McLerran:2018hbz, Zhao:2020dvu}. Building on these studies, we consider a general power-law parametrization of the momentum shell width:
\begin{equation}\label{eq_shell}
\Delta = \Lambda_{\rm QCD} \left(\frac{\Lambda_{\rm QCD}}{k_F}\right)^\alpha \textrm,
\end{equation}
where $k_F$ is the Fermi momentum corresponding to the nucleon, and $\alpha$ is a dimensionless parameter controlling how rapidly the shell width decreases with increasing density. This parametrization incorporates previous studies as special cases. For example, $\alpha = 1$ recovers earlier work in Ref.~\cite{Zhao:2020dvu} and $\alpha = 2$ corresponds to the form used in Ref.~\cite{McLerran:2018hbz}. The case $\alpha = 0$ represents a constant shell width $\Delta = \Lambda_{\rm QCD}$. In this work, we fix $\Lambda_{\rm QCD} = 300~$MeV. Treating $\alpha$ as a constant, the in-medium profiles of the mean fields are obtained through the gap equations,
\begin{equation}
\frac{\partial \Omega}{\partial \sigma} = 0, \quad \frac{\partial \Omega}{\partial \omega} = 0 \textrm.
\label{eq_gap}
\end{equation}
Using the parametrization in Eq.~\eqref{eq_shell}, one expresses the baryon number density purely as a function of $k_F$ in both quarkyonic and baryquark scenarios. We note that to maintain consistency between the quarkyonic and baryquark descriptions, we fix the total Fermi momentum $k_F$ in both scenarios.

In Fig.~\ref{fig_nb_kf}, we show the baryon number density as a function of the Fermi momentum $k_F$ for different choices of $\alpha$. For $\alpha=0$, the momentum shell $\Delta$ is independent of $k_F$ and the baryon density increases monotonically in both models, even beyond the point where $k_F > \Delta$. The behavior changes for $\alpha>0$. The baryon number density decreases after the transition to the quarkyonic or baryquark phase. This happens for $\alpha \gtrsim 3$ in the quarkyonic model and for $\alpha \gtrsim 0.1$ in the baryquark model. This counterintuitive behavior can be understood from the $N_c^{-3}$ suppression of the quark contribution to the baryon density. In the baryquark model, the quark fraction $Y_q \equiv n_q / n_B$ increases more rapidly with density compared to the quarkyonic phase, as shown in the bottom panel of Fig.~\ref{fig_nb_kf}. Consequently, the suppression of the baryon density becomes particularly significant, i.e., the quark fraction grows too rapidly, which leads to an overall decrease in the total baryon density. At sufficiently large Fermi momentum $k_F$, baryon density starts to increase again. Clearly, thermodynamic consistency yields maximum values of $\alpha$, which showcases the downside of the ansatz in Eq.~\eqref{eq_shell}. To overcome this problem, we treat the momentum-space shell self-consistently in the next section.

\section{Dynamically generated momentum-space shell}
\label{sec:delta2}
In this section, we do not assume any functional form of the momentum shell $\Delta$ and allow it to be a free parameter. That leaves us with three variables to determine. We proceed by fixing the total baryon density $n_B$, which fixes the value of the $\omega$ mean field (cf. Eq.~\eqref{eq_gap}). Then, we allow the quark fraction $Y_q$ to vary and fix the momentum shell by minimizing the energy density~\cite{Jeong:2019lhv, Koch:2022act}. We note that minimization of the energy density at fixed baryon density is equivalent to solving an additional gap equation, i.e.,
\begin{equation}\label{eq:yq_gap_eq}
\frac{\pp \eps}{\pp Y_q}\Bigg|_{n_B} = 0 \Longleftrightarrow \frac{\pp \Omega}{\pp Y_q}\Bigg|_{n_B} = 0.
\end{equation}

We examine two scenarios for quark interactions. First, where quarks couple to the $\sigma$ meson, i.e., $m_q = m_N/3$; second, where quarks are constrained only by the Pauli principle and behave as free particles of constant mass fixed to the vacuum value, i.e., $m_q = m_N^{\rm vaq}/3$ [cf.~Eq.~\eqref{eq:quark_mass}]. 

\begin{figure}[!t]
\includegraphics[width=1\hsize]{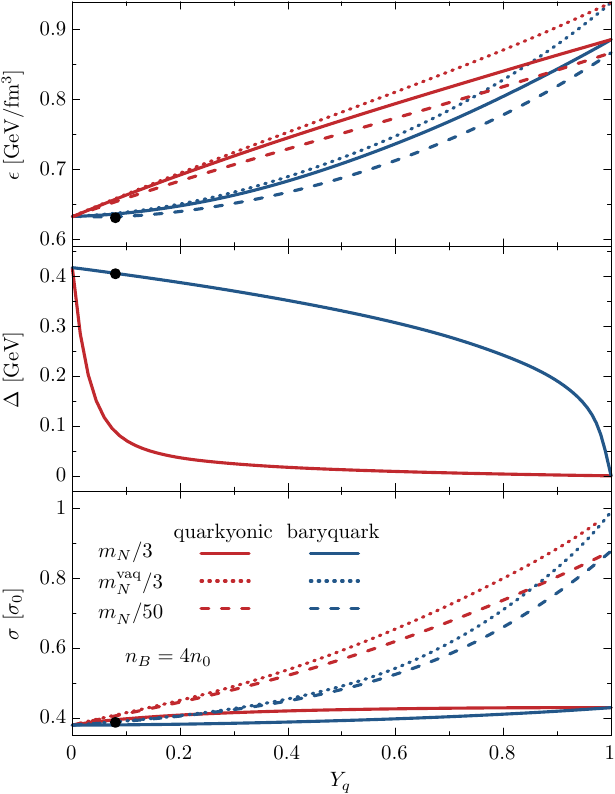}
\caption{Energy density (top panel), momentum shell (middle panel), and the expectation value of the $\sigma$ mean field (bottom panel) for different parametrizations of the quark mass, as functions of quark fraction $Y_q$ for fixed baryon density $n_B=4n_0$. In all panels, black dots mark the onset of baryquark phase for $m_q = m_N/{50}$ (see text for details).}
\label{fig:yq_panel}
\end{figure}

Fig.~\ref{fig:yq_panel} illustrates these scenarios at fixed baryon density $n_B = 4n_0$ as an example. In the middle panel, we show the momentum shell structure in the baryquark and quarkyonic models. The momentum shell decreases rapidly in the quarkyonic model. On the other hand, in the baryquark model, the decrease is slow and almost linear up to $Y_q \simeq 0.7$. After that, $\Delta$ quickly vanishes as $Y_q$ goes to unity. Interestingly, within the model, the momentum shells are identical in the quarkyonic and baryquark scenarios across all cases considered. This is because the value of $\Delta$ can be derived solely from the Fermi momentum $k_F^3 = 3\pi^2n_B/2$, which is the same at fixed $n_B$ and $Y_q$. 

\begin{figure}\centering
\includegraphics[width=1\hsize]{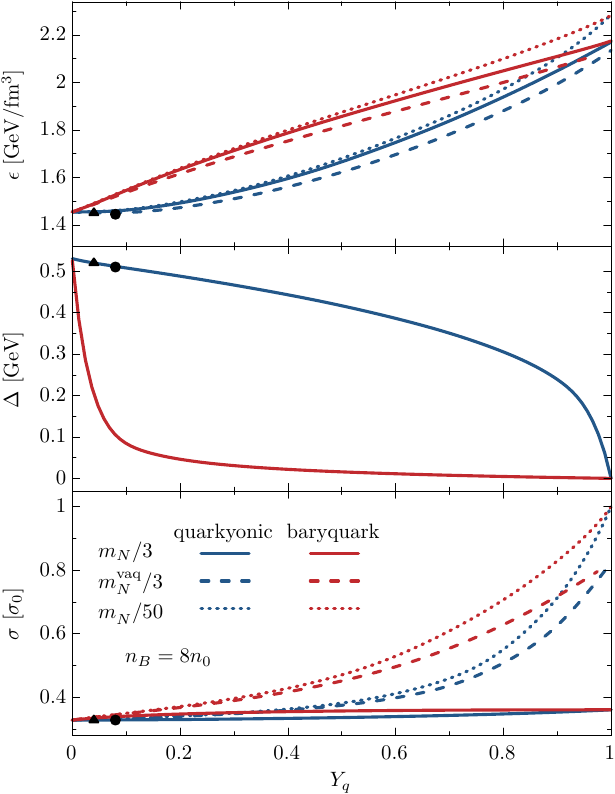}
\caption{As in Fig.~\ref{fig:yq_panel} but at $n_B=8n_0$. Triangles mark the onset of baryquark phase for the physical case with $m_q = m_N/3$ (see text for details).}
\label{fig:energy_8n0}
\end{figure}

Due to the gap equation, the vector mean field $\omega \propto n_B$; thus, one expects that at fixed density $n_B$ the variation of the quark fraction does not change the expectation value of the $\omega$ mean field. Consequently, the change of quark fraction at fixed baryon density can only be accommodated by the change of the $\sigma$ mean field, and thus, in the constituent masses of nucleons and/or quarks. This is seen in the bottom panel of Fig.~\ref{fig:yq_panel}, which depicts the $\sigma$ mean field value as a function of quark fraction $Y_q$. In the free quark scenario ($m_q=m_N^{\rm vac}/3$), the $\sigma$ expectation value increases as the quark fraction goes from zero to unity and reaches its vacuum value at $Y_q=1$. This behavior is understood since the nucleon density becomes suppressed with increasing quark fraction. Consequently, the chiral symmetry breaking is driven because quarks are not coupled to the $\sigma$ field. Therefore, $\sigma$ naturally returns to its vacuum value. In contrast, when quarks couple to the $\sigma$ mean field ($m_q=m_N/3$), its value is almost constant, and increases only slightly with increasing quark fraction $Y_q$, indicating persistent chiral symmetry breaking for all values of $Y_q$. 

The behavior of the energy density at finite quark fraction is similar. It is depicted in the top panel of Fig.~\ref{fig:yq_panel}. The energy density increases with increasing quark fraction in both scenarios. When quarks couple to the $\sigma$ mean field, both nucleon and constituent quark masses decrease while an additional attractive force contributes negatively to the potential, resulting in lower energy density, compared to the free quark scenario. We also find that the energy density is always smaller in the baryquark model. This shows that the self-consistent treatment of the momentum shell renders baryquark matter energetically favored over quarkyonic matter. 

A qualitatively similar result was found in a model with constituent quarks and hard-core repulsion with an assumption that the chiral symmetry is broken~\cite{Koch:2022act}. Albeit the baryquark matter is energetically favorable at finite $Y_q$, we find that in both models the energy density is minimized for $Y_q = 0$, thus dismissing the onset of quarks in either of the models, leaving the matter purely hadronic. We have verified that the results are qualitatively the same for different values of the chirally invariant mass $m_0$. 

We have to note that there are several important differences that distinguish our approach from that of Ref.~\cite{Koch:2022act}. In their approach, nucleon repulsive forces arise from excluded volume effects among hadrons, which gradually diminish as the nucleon density decreases. Consequently, when $Y_q=1$, the excluded volume model reduces to a system of free quarks with neither attractive nor repulsive interactions. Our formalism incorporates both attractive and repulsive forces through a self-consistent mean-field treatment. The $\omega$ mean field remains fixed at given baryon density by means of the gap equation $\partial\Omega/\partial\omega = 0$. This means the repulsive $\omega$ contribution persists even in the pure quark phase at $Y_q=1$, leading to an increased energy density of the whole system compared to the free quark scenario. Furthermore, the presence of the attractive $\sigma$ interaction significantly delays the onset of quark degrees of freedom, pushing their appearance to much higher densities than in the excluded volume model. Therefore, the emergence of quarks at moderate densities is energetically unfavorable in our approach, in contrast to models based purely on geometric excluded volume considerations. 

To achieve the onset of quark degrees of freedom, i.e., a minimum of energy density at finite quark fraction, we explore several approaches. One way is to simply increase the baryon density. We find that in the baryquark model, the onset of quark degrees of freedom takes place around $n_B \simeq 8n_0$. This is shown in Fig.~\ref{fig:energy_8n0}. At $n_B=8n_0$, the energy density attains its minimum at a finite value of quark fraction $Y_q\simeq 0.04$. On the other hand, purely hadronic matter is still favored in the quarkyonic model. In general, the chiral condensate exhibits only weak sensitivity to finite quark fractions at lower density. It drives the chiral symmetry breaking as the quark content increases, but tends towards chiral symmetry restoration os the baryon density increases. This mechanism explains why the onset of quarks is delayed even when partial chiral restoration is already underway. However, we note that $n_B=8n_0$ corresponds to $\mu_B\simeq 1.5~\rm GeV$, which is at the border of the applicability of the current model, i.e., roughly where the onset of the negative-parity chiral partner of the nucleon is expected. This suggests that the onset of quark degrees of freedom is not favored in the current model of quarkyonic and baryquark matter.

Another way involves a modification of the constituent quark mass. To exemplify this, we set $m_q = m_N/50$. We note that this is equivalent to almost massless quarks. The corresponding energy-density profiles at $n_B=4n_0$ and $8n_0$ are shown in Fig.~\ref{fig:yq_panel} and Fig.~\ref{fig:energy_8n0}, respectively. In both quarkyonic and baryquark models, the energy density is smaller compared to the physical case with $m_q=m_N/3$. However, only in the case of baryquark matter, a minimum is achieved at $Y_q\simeq 0.08$ at $n_B = 4n_0$. At $n_B=8n_0$, the onset of quark degrees of freedom is achieved at roughly the same value of $Y_q \simeq 0.08$. This means that the quark fraction is almost constant starting from the onset of baryquark matter to several times the saturation density (i.e., up to the applicability of the model). The modification of the quark mass has notable effects on the $\sigma$ mean field, which increases much more rapidly. This behavior arises because the quark contribution to the gap equation becomes negligible for extremely small $m_q$. In the limit $m_q \rightarrow 0$, quarks effectively decouple from the chiral dynamics, leaving the gap equation governed almost entirely by the nucleon sector. Consequently, the $\sigma$ field tends to return toward its vacuum value, similar to the scenario where quarks are treated as non-interacting free particles that do not couple to $\sigma$. Interestingly, while both massless constituent quarks and non-interacting quarks exhibit similar $\sigma$ field evolution, they produce qualitatively different energy profiles. In the case where quarks do not couple to the $\sigma$ mean field, they behave as free particles which do not generate any attractive interactions, resulting in a relatively higher energy density. In contrast, when quarks become nearly massless but remain coupled to the $\sigma$ field, they induce a dynamical attractive force that lowers the total energy density of the system. This finding aligns with the results in Ref.~\cite{Gao:2024jlp}, which showed that the density dependence of constituent quark masses stiffens the equation of state (reducing the energy density). These results have important implications for the relationship between quark degrees of freedom and chiral symmetry. Notably, our findings suggest that the onset of quark degrees of freedom need not coincide with chiral symmetry restoration.

\section{Summary}
\label{sec:summary}

In this work, we have studied the role of quark excitations in dense nuclear matter. We have formulated and compared two schematic constructions for baryonic matter with explicit quark degrees of freedom: a quarkyonic model, where quarks occupy momentum shells outside the baryonic Fermi sea, and a baryquark model, where quarks appear inside the Fermi surface. Both realizations were embedded in a parity-doublet mean-field model to provide a consistent description of chiral symmetry. 

Our results show that a self-consistent treatment of the shell structure makes baryquark matter energetically favored relative to quarkyonic matter. However, both remain disfavored compared to purely hadronic matter up to several times nuclear saturation density. The hadronic equation of state continues to dominate, and quark degrees of freedom do not play a significant role within the model’s applicability range. Importantly, we demonstrate that the onset of quark matter does not necessarily coincide with the restoration of chiral symmetry, indicating that the two phenomena may be more weakly correlated than commonly assumed.

These findings suggest that the location and character of quark deconfinement depend sensitively on how quarks are embedded relative to the baryonic Fermi surface. While schematic, the present framework establishes a controlled setting to explore competing quark–baryon scenarios and their interplay with chiral symmetry. 

Future work should explore extensions of the present framework by incorporating additional quark interactions, such as diquark pairing or color superconductivity, as well as a more realistic treatment of confinement. These mechanisms may shift the balance in favor of quark degrees of freedom and could provide a more complete picture of dense matter in QCD. Generalization to finite isospin asymmetry and direct confrontation with neutron-star phenomenology will be important directions for future work.

\medskip
\acknowledgments
 M.~M. acknowledges the support through the program Excellence Initiative–Research University of the University of Wroc\l{}aw of the Ministry of Education and Science.
\bibliography{ref_quarkyonic}

\begin{thebibliography}{64}%
\makeatletter
\providecommand \@ifxundefined [1]{%
 \@ifx{#1\undefined}
}%
\providecommand \@ifnum [1]{%
 \ifnum #1\expandafter \@firstoftwo
 \else \expandafter \@secondoftwo
 \fi
}%
\providecommand \@ifx [1]{%
 \ifx #1\expandafter \@firstoftwo
 \else \expandafter \@secondoftwo
 \fi
}%
\providecommand \natexlab [1]{#1}%
\providecommand \enquote  [1]{``#1''}%
\providecommand \bibnamefont  [1]{#1}%
\providecommand \bibfnamefont [1]{#1}%
\providecommand \citenamefont [1]{#1}%
\providecommand \href@noop [0]{\@secondoftwo}%
\providecommand \href [0]{\begingroup \@sanitize@url \@href}%
\providecommand \@href[1]{\@@startlink{#1}\@@href}%
\providecommand \@@href[1]{\endgroup#1\@@endlink}%
\providecommand \@sanitize@url [0]{\catcode `\\12\catcode `\$12\catcode
  `\&12\catcode `\#12\catcode `\^12\catcode `\_12\catcode `\%12\relax}%
\providecommand \@@startlink[1]{}%
\providecommand \@@endlink[0]{}%
\providecommand \url  [0]{\begingroup\@sanitize@url \@url }%
\providecommand \@url [1]{\endgroup\@href {#1}{\urlprefix }}%
\providecommand \urlprefix  [0]{URL }%
\providecommand \Eprint [0]{\href }%
\providecommand \doibase [0]{https://doi.org/}%
\providecommand \selectlanguage [0]{\@gobble}%
\providecommand \bibinfo  [0]{\@secondoftwo}%
\providecommand \bibfield  [0]{\@secondoftwo}%
\providecommand \translation [1]{[#1]}%
\providecommand \BibitemOpen [0]{}%
\providecommand \bibitemStop [0]{}%
\providecommand \bibitemNoStop [0]{.\EOS\space}%
\providecommand \EOS [0]{\spacefactor3000\relax}%
\providecommand \BibitemShut  [1]{\csname bibitem#1\endcsname}%
\let\auto@bib@innerbib\@empty
\bibitem [{\citenamefont {Miller}\ \emph {et~al.}(2021)\citenamefont {Miller}
  \emph {et~al.}}]{Miller:2021qha}%
  \BibitemOpen
  \bibfield  {author} {\bibinfo {author} {\bibfnamefont {M.~C.}\ \bibnamefont
  {Miller}} \emph {et~al.},\ }\bibfield  {title} {\bibinfo {title} {{The Radius
  of PSR J0740+6620 from NICER and XMM-Newton Data}},\ }\href
  {https://doi.org/10.3847/2041-8213/ac089b} {\bibfield  {journal} {\bibinfo
  {journal} {Astrophys. J. Lett.}\ }\textbf {\bibinfo {volume} {918}},\
  \bibinfo {pages} {L28} (\bibinfo {year} {2021})},\ \Eprint
  {https://arxiv.org/abs/2105.06979} {arXiv:2105.06979 [astro-ph.HE]}
  \BibitemShut {NoStop}%
\bibitem [{\citenamefont {Riley}\ \emph {et~al.}(2021)\citenamefont {Riley},
  \citenamefont {Watts}, \citenamefont {Ray}, \citenamefont {Bogdanov},
  \citenamefont {Guillot}, \citenamefont {Morsink}, \citenamefont {Bilous},
  \citenamefont {Arzoumanian}, \citenamefont {Choudhury}, \citenamefont
  {Deneva}, \citenamefont {Gendreau}, \citenamefont {Harding}, \citenamefont
  {Ho}, \citenamefont {Lattimer}, \citenamefont {Loewenstein}, \citenamefont
  {Ludlam}, \citenamefont {Markwardt}, \citenamefont {Okajima}, \citenamefont
  {Prescod-Weinstein}, \citenamefont {Remillard}, \citenamefont {Wolff},
  \citenamefont {Fonseca}, \citenamefont {Cromartie}, \citenamefont {Kerr},
  \citenamefont {Pennucci}, \citenamefont {Parthasarathy}, \citenamefont
  {Ransom}, \citenamefont {Stairs}, \citenamefont {Guillemot},\ and\
  \citenamefont {Cognard}}]{Riley:2021pdl}%
  \BibitemOpen
  \bibfield  {author} {\bibinfo {author} {\bibfnamefont {T.~E.}\ \bibnamefont
  {Riley}}, \bibinfo {author} {\bibfnamefont {A.~L.}\ \bibnamefont {Watts}},
  \bibinfo {author} {\bibfnamefont {P.~S.}\ \bibnamefont {Ray}}, \bibinfo
  {author} {\bibfnamefont {S.}~\bibnamefont {Bogdanov}}, \bibinfo {author}
  {\bibfnamefont {S.}~\bibnamefont {Guillot}}, \bibinfo {author} {\bibfnamefont
  {S.~M.}\ \bibnamefont {Morsink}}, \bibinfo {author} {\bibfnamefont {A.~V.}\
  \bibnamefont {Bilous}}, \bibinfo {author} {\bibfnamefont {Z.}~\bibnamefont
  {Arzoumanian}}, \bibinfo {author} {\bibfnamefont {D.}~\bibnamefont
  {Choudhury}}, \bibinfo {author} {\bibfnamefont {J.~S.}\ \bibnamefont
  {Deneva}}, \bibinfo {author} {\bibfnamefont {K.~C.}\ \bibnamefont
  {Gendreau}}, \bibinfo {author} {\bibfnamefont {A.~K.}\ \bibnamefont
  {Harding}}, \bibinfo {author} {\bibfnamefont {W.~C.~G.}\ \bibnamefont {Ho}},
  \bibinfo {author} {\bibfnamefont {J.~M.}\ \bibnamefont {Lattimer}}, \bibinfo
  {author} {\bibfnamefont {M.}~\bibnamefont {Loewenstein}}, \bibinfo {author}
  {\bibfnamefont {R.~M.}\ \bibnamefont {Ludlam}}, \bibinfo {author}
  {\bibfnamefont {C.~B.}\ \bibnamefont {Markwardt}}, \bibinfo {author}
  {\bibfnamefont {T.}~\bibnamefont {Okajima}}, \bibinfo {author} {\bibfnamefont
  {C.}~\bibnamefont {Prescod-Weinstein}}, \bibinfo {author} {\bibfnamefont
  {R.~A.}\ \bibnamefont {Remillard}}, \bibinfo {author} {\bibfnamefont {M.~T.}\
  \bibnamefont {Wolff}}, \bibinfo {author} {\bibfnamefont {E.}~\bibnamefont
  {Fonseca}}, \bibinfo {author} {\bibfnamefont {H.~T.}\ \bibnamefont
  {Cromartie}}, \bibinfo {author} {\bibfnamefont {M.}~\bibnamefont {Kerr}},
  \bibinfo {author} {\bibfnamefont {T.~T.}\ \bibnamefont {Pennucci}}, \bibinfo
  {author} {\bibfnamefont {A.}~\bibnamefont {Parthasarathy}}, \bibinfo {author}
  {\bibfnamefont {S.}~\bibnamefont {Ransom}}, \bibinfo {author} {\bibfnamefont
  {I.}~\bibnamefont {Stairs}}, \bibinfo {author} {\bibfnamefont
  {L.}~\bibnamefont {Guillemot}},\ and\ \bibinfo {author} {\bibfnamefont
  {I.}~\bibnamefont {Cognard}},\ }\bibfield  {title} {\bibinfo {title} {A nicer
  view of the massive pulsar psr j0740+6620 informed by radio timing and
  xmm-newton spectroscopy},\ }\href {https://doi.org/10.3847/2041-8213/ac0a81}
  {\bibfield  {journal} {\bibinfo  {journal} {The Astrophysical Journal
  Letters}\ }\textbf {\bibinfo {volume} {918}},\ \bibinfo {pages} {L27}
  (\bibinfo {year} {2021})}\BibitemShut {NoStop}%
\bibitem [{\citenamefont {Abbott}\ \emph
  {et~al.}(2017{\natexlab{a}})\citenamefont {Abbott} \emph
  {et~al.}}]{LIGOScientific:2017vwq}%
  \BibitemOpen
  \bibfield  {author} {\bibinfo {author} {\bibfnamefont {B.~P.}\ \bibnamefont
  {Abbott}} \emph {et~al.} (\bibinfo {collaboration} {LIGO Scientific,
  Virgo}),\ }\bibfield  {title} {\bibinfo {title} {{GW170817: Observation of
  Gravitational Waves from a Binary Neutron Star Inspiral}},\ }\href
  {https://doi.org/10.1103/PhysRevLett.119.161101} {\bibfield  {journal}
  {\bibinfo  {journal} {Phys. Rev. Lett.}\ }\textbf {\bibinfo {volume} {119}},\
  \bibinfo {pages} {161101} (\bibinfo {year} {2017}{\natexlab{a}})},\ \Eprint
  {https://arxiv.org/abs/1710.05832} {arXiv:1710.05832 [gr-qc]} \BibitemShut
  {NoStop}%
\bibitem [{\citenamefont {Abbott}\ \emph
  {et~al.}(2017{\natexlab{b}})\citenamefont {Abbott} \emph
  {et~al.}}]{LIGOScientific:2017ync}%
  \BibitemOpen
  \bibfield  {author} {\bibinfo {author} {\bibfnamefont {B.~P.}\ \bibnamefont
  {Abbott}} \emph {et~al.} (\bibinfo {collaboration} {LIGO Scientific, Virgo,
  Fermi GBM, INTEGRAL, IceCube, AstroSat Cadmium Zinc Telluride Imager Team,
  IPN, Insight-Hxmt, ANTARES, Swift, AGILE Team, 1M2H Team, Dark Energy Camera
  GW-EM, DES, DLT40, GRAWITA, Fermi-LAT, ATCA, ASKAP, Las Cumbres Observatory
  Group, OzGrav, DWF (Deeper Wider Faster Program), AST3, CAASTRO, VINROUGE,
  MASTER, J-GEM, GROWTH, JAGWAR, CaltechNRAO, TTU-NRAO, NuSTAR, Pan-STARRS,
  MAXI Team, TZAC Consortium, KU, Nordic Optical Telescope, ePESSTO, GROND,
  Texas Tech University, SALT Group, TOROS, BOOTES, MWA, CALET, IKI-GW
  Follow-up, H.E.S.S., LOFAR, LWA, HAWC, Pierre Auger, ALMA, Euro VLBI Team, Pi
  of Sky, Chandra Team at McGill University, DFN, ATLAS Telescopes, High Time
  Resolution Universe Survey, RIMAS, RATIR, SKA South Africa/MeerKAT}),\
  }\bibfield  {title} {\bibinfo {title} {{Multi-messenger Observations of a
  Binary Neutron Star Merger}},\ }\href
  {https://doi.org/10.3847/2041-8213/aa91c9} {\bibfield  {journal} {\bibinfo
  {journal} {Astrophys. J. Lett.}\ }\textbf {\bibinfo {volume} {848}},\
  \bibinfo {pages} {L12} (\bibinfo {year} {2017}{\natexlab{b}})},\ \Eprint
  {https://arxiv.org/abs/1710.05833} {arXiv:1710.05833 [astro-ph.HE]}
  \BibitemShut {NoStop}%
\bibitem [{\citenamefont {Lenzi}\ and\ \citenamefont
  {Lugones}(2012)}]{Lenzi:2012xz}%
  \BibitemOpen
  \bibfield  {author} {\bibinfo {author} {\bibfnamefont {C.~H.}\ \bibnamefont
  {Lenzi}}\ and\ \bibinfo {author} {\bibfnamefont {G.}~\bibnamefont
  {Lugones}},\ }\bibfield  {title} {\bibinfo {title} {{Hybrid stars in the
  light of the massive pulsar PSR J1614-2230}},\ }\href
  {https://doi.org/10.1088/0004-637X/759/1/57} {\bibfield  {journal} {\bibinfo
  {journal} {Astrophys. J.}\ }\textbf {\bibinfo {volume} {759}},\ \bibinfo
  {pages} {57} (\bibinfo {year} {2012})},\ \Eprint
  {https://arxiv.org/abs/1206.4108} {arXiv:1206.4108 [astro-ph.SR]}
  \BibitemShut {NoStop}%
\bibitem [{\citenamefont {Benic}\ \emph {et~al.}(2015)\citenamefont {Benic},
  \citenamefont {Blaschke}, \citenamefont {Alvarez-Castillo}, \citenamefont
  {Fischer},\ and\ \citenamefont {Typel}}]{Benic:2014jia}%
  \BibitemOpen
  \bibfield  {author} {\bibinfo {author} {\bibfnamefont {S.}~\bibnamefont
  {Benic}}, \bibinfo {author} {\bibfnamefont {D.}~\bibnamefont {Blaschke}},
  \bibinfo {author} {\bibfnamefont {D.~E.}\ \bibnamefont {Alvarez-Castillo}},
  \bibinfo {author} {\bibfnamefont {T.}~\bibnamefont {Fischer}},\ and\ \bibinfo
  {author} {\bibfnamefont {S.}~\bibnamefont {Typel}},\ }\bibfield  {title}
  {\bibinfo {title} {{A new quark-hadron hybrid equation of state for
  astrophysics - I. High-mass twin compact stars}},\ }\href
  {https://doi.org/10.1051/0004-6361/201425318} {\bibfield  {journal} {\bibinfo
   {journal} {Astron. Astrophys.}\ }\textbf {\bibinfo {volume} {577}},\
  \bibinfo {pages} {A40} (\bibinfo {year} {2015})},\ \Eprint
  {https://arxiv.org/abs/1411.2856} {arXiv:1411.2856 [astro-ph.HE]}
  \BibitemShut {NoStop}%
\bibitem [{\citenamefont {G\"artlein}\ \emph {et~al.}(2023)\citenamefont
  {G\"artlein}, \citenamefont {Ivanytskyi}, \citenamefont {Sagun},\ and\
  \citenamefont {Blaschke}}]{Gartlein:2023vif}%
  \BibitemOpen
  \bibfield  {author} {\bibinfo {author} {\bibfnamefont {C.}~\bibnamefont
  {G\"artlein}}, \bibinfo {author} {\bibfnamefont {O.}~\bibnamefont
  {Ivanytskyi}}, \bibinfo {author} {\bibfnamefont {V.}~\bibnamefont {Sagun}},\
  and\ \bibinfo {author} {\bibfnamefont {D.}~\bibnamefont {Blaschke}},\
  }\bibfield  {title} {\bibinfo {title} {{Hybrid star phenomenology from the
  properties of the special point}},\ }\href
  {https://doi.org/10.1103/PhysRevD.108.114028} {\bibfield  {journal} {\bibinfo
   {journal} {Phys. Rev. D}\ }\textbf {\bibinfo {volume} {108}},\ \bibinfo
  {pages} {114028} (\bibinfo {year} {2023})},\ \Eprint
  {https://arxiv.org/abs/2301.10765} {arXiv:2301.10765 [nucl-th]} \BibitemShut
  {NoStop}%
\bibitem [{\citenamefont {Carlomagno}\ \emph {et~al.}(2024)\citenamefont
  {Carlomagno}, \citenamefont {Contrera}, \citenamefont {Grunfeld},\ and\
  \citenamefont {Blaschke}}]{Carlomagno:2023nrc}%
  \BibitemOpen
  \bibfield  {author} {\bibinfo {author} {\bibfnamefont {J.~P.}\ \bibnamefont
  {Carlomagno}}, \bibinfo {author} {\bibfnamefont {G.~A.}\ \bibnamefont
  {Contrera}}, \bibinfo {author} {\bibfnamefont {A.~G.}\ \bibnamefont
  {Grunfeld}},\ and\ \bibinfo {author} {\bibfnamefont {D.}~\bibnamefont
  {Blaschke}},\ }\bibfield  {title} {\bibinfo {title} {{Thermal twin stars
  within a hybrid equation of state based on a nonlocal chiral quark model
  compatible with modern astrophysical observations}},\ }\href
  {https://doi.org/10.1103/PhysRevD.109.043050} {\bibfield  {journal} {\bibinfo
   {journal} {Phys. Rev. D}\ }\textbf {\bibinfo {volume} {109}},\ \bibinfo
  {pages} {043050} (\bibinfo {year} {2024})},\ \Eprint
  {https://arxiv.org/abs/2312.01975} {arXiv:2312.01975 [nucl-th]} \BibitemShut
  {NoStop}%
\bibitem [{\citenamefont {Ayriyan}\ \emph {et~al.}(2025)\citenamefont
  {Ayriyan}, \citenamefont {Blaschke}, \citenamefont {Carlomagno},
  \citenamefont {Contrera},\ and\ \citenamefont {Grunfeld}}]{Ayriyan:2024zfw}%
  \BibitemOpen
  \bibfield  {author} {\bibinfo {author} {\bibfnamefont {A.}~\bibnamefont
  {Ayriyan}}, \bibinfo {author} {\bibfnamefont {D.}~\bibnamefont {Blaschke}},
  \bibinfo {author} {\bibfnamefont {J.~P.}\ \bibnamefont {Carlomagno}},
  \bibinfo {author} {\bibfnamefont {G.~A.}\ \bibnamefont {Contrera}},\ and\
  \bibinfo {author} {\bibfnamefont {A.~G.}\ \bibnamefont {Grunfeld}},\
  }\bibfield  {title} {\bibinfo {title} {{Bayesian Analysis of Hybrid Neutron
  Star EOS Constraints Within an Instantaneous Nonlocal Chiral Quark Matter
  Model}},\ }\href {https://doi.org/10.3390/universe11050141} {\bibfield
  {journal} {\bibinfo  {journal} {Universe}\ }\textbf {\bibinfo {volume}
  {11}},\ \bibinfo {pages} {141} (\bibinfo {year} {2025})},\ \Eprint
  {https://arxiv.org/abs/2501.00115} {arXiv:2501.00115 [nucl-th]} \BibitemShut
  {NoStop}%
\bibitem [{\citenamefont {Gao}\ \emph {et~al.}(2024{\natexlab{a}})\citenamefont
  {Gao}, \citenamefont {Yuan}, \citenamefont {Harada},\ and\ \citenamefont
  {Ma}}]{Gao:2024lzu}%
  \BibitemOpen
  \bibfield  {author} {\bibinfo {author} {\bibfnamefont {B.}~\bibnamefont
  {Gao}}, \bibinfo {author} {\bibfnamefont {W.-L.}\ \bibnamefont {Yuan}},
  \bibinfo {author} {\bibfnamefont {M.}~\bibnamefont {Harada}},\ and\ \bibinfo
  {author} {\bibfnamefont {Y.-L.}\ \bibnamefont {Ma}},\ }\bibfield  {title}
  {\bibinfo {title} {{Exploring the first-order phase transition in neutron
  stars using the parity doublet model and a
  Nambu\textendash{}Jona-Lasinio\textendash{}type quark model}},\ }\href
  {https://doi.org/10.1103/PhysRevC.110.045802} {\bibfield  {journal} {\bibinfo
   {journal} {Phys. Rev. C}\ }\textbf {\bibinfo {volume} {110}},\ \bibinfo
  {pages} {045802} (\bibinfo {year} {2024}{\natexlab{a}})},\ \Eprint
  {https://arxiv.org/abs/2407.13990} {arXiv:2407.13990 [nucl-th]} \BibitemShut
  {NoStop}%
\bibitem [{\citenamefont {Christian}\ \emph {et~al.}(2025)\citenamefont
  {Christian}, \citenamefont {Rather}, \citenamefont {Gholami},\ and\
  \citenamefont {Hofmann}}]{Christian:2025dhe}%
  \BibitemOpen
  \bibfield  {author} {\bibinfo {author} {\bibfnamefont {J.-E.}\ \bibnamefont
  {Christian}}, \bibinfo {author} {\bibfnamefont {I.~A.}\ \bibnamefont
  {Rather}}, \bibinfo {author} {\bibfnamefont {H.}~\bibnamefont {Gholami}},\
  and\ \bibinfo {author} {\bibfnamefont {M.}~\bibnamefont {Hofmann}},\
  }\href@noop {} {\bibinfo {title} {{Comprehensive Analysis of Constructing
  Hybrid Stars with an RG-consistent NJL Model}}} (\bibinfo {year} {2025}),\
  \Eprint {https://arxiv.org/abs/2503.13626} {arXiv:2503.13626 [astro-ph.HE]}
  \BibitemShut {NoStop}%
\bibitem [{\citenamefont {Yang}\ \emph {et~al.}(2025)\citenamefont {Yang},
  \citenamefont {Zeng}, \citenamefont {Yan}, \citenamefont {Yuan},
  \citenamefont {Zhang},\ and\ \citenamefont {Zhou}}]{Yang:2025iyv}%
  \BibitemOpen
  \bibfield  {author} {\bibinfo {author} {\bibfnamefont {Z.}~\bibnamefont
  {Yang}}, \bibinfo {author} {\bibfnamefont {T.}~\bibnamefont {Zeng}}, \bibinfo
  {author} {\bibfnamefont {Y.}~\bibnamefont {Yan}}, \bibinfo {author}
  {\bibfnamefont {W.-L.}\ \bibnamefont {Yuan}}, \bibinfo {author}
  {\bibfnamefont {C.}~\bibnamefont {Zhang}},\ and\ \bibinfo {author}
  {\bibfnamefont {E.}~\bibnamefont {Zhou}},\ }\href@noop {} {\bibinfo {title}
  {{Hybrid Quark Stars with Quark-Quark Phase Transitions}}} (\bibinfo {year}
  {2025}),\ \Eprint {https://arxiv.org/abs/2507.00776} {arXiv:2507.00776
  [astro-ph.HE]} \BibitemShut {NoStop}%
\bibitem [{\citenamefont {Masuda}\ \emph
  {et~al.}(2013{\natexlab{a}})\citenamefont {Masuda}, \citenamefont {Hatsuda},\
  and\ \citenamefont {Takatsuka}}]{Masuda:2012kf}%
  \BibitemOpen
  \bibfield  {author} {\bibinfo {author} {\bibfnamefont {K.}~\bibnamefont
  {Masuda}}, \bibinfo {author} {\bibfnamefont {T.}~\bibnamefont {Hatsuda}},\
  and\ \bibinfo {author} {\bibfnamefont {T.}~\bibnamefont {Takatsuka}},\
  }\bibfield  {title} {\bibinfo {title} {{Hadron-Quark Crossover and Massive
  Hybrid Stars with Strangeness}},\ }\href
  {https://doi.org/10.1088/0004-637X/764/1/12} {\bibfield  {journal} {\bibinfo
  {journal} {Astrophys. J.}\ }\textbf {\bibinfo {volume} {764}},\ \bibinfo
  {pages} {12} (\bibinfo {year} {2013}{\natexlab{a}})},\ \Eprint
  {https://arxiv.org/abs/1205.3621} {arXiv:1205.3621 [nucl-th]} \BibitemShut
  {NoStop}%
\bibitem [{\citenamefont {Masuda}\ \emph
  {et~al.}(2013{\natexlab{b}})\citenamefont {Masuda}, \citenamefont {Hatsuda},\
  and\ \citenamefont {Takatsuka}}]{Masuda:2012ed}%
  \BibitemOpen
  \bibfield  {author} {\bibinfo {author} {\bibfnamefont {K.}~\bibnamefont
  {Masuda}}, \bibinfo {author} {\bibfnamefont {T.}~\bibnamefont {Hatsuda}},\
  and\ \bibinfo {author} {\bibfnamefont {T.}~\bibnamefont {Takatsuka}},\
  }\bibfield  {title} {\bibinfo {title} {{Hadron{\textendash}quark crossover
  and massive hybrid stars}},\ }\href {https://doi.org/10.1093/ptep/ptt045}
  {\bibfield  {journal} {\bibinfo  {journal} {PTEP}\ }\textbf {\bibinfo
  {volume} {2013}},\ \bibinfo {pages} {073D01} (\bibinfo {year}
  {2013}{\natexlab{b}})},\ \Eprint {https://arxiv.org/abs/1212.6803}
  {arXiv:1212.6803 [nucl-th]} \BibitemShut {NoStop}%
\bibitem [{\citenamefont {Kojo}\ \emph {et~al.}(2015)\citenamefont {Kojo},
  \citenamefont {Powell}, \citenamefont {Song},\ and\ \citenamefont
  {Baym}}]{Kojo:2014rca}%
  \BibitemOpen
  \bibfield  {author} {\bibinfo {author} {\bibfnamefont {T.}~\bibnamefont
  {Kojo}}, \bibinfo {author} {\bibfnamefont {P.~D.}\ \bibnamefont {Powell}},
  \bibinfo {author} {\bibfnamefont {Y.}~\bibnamefont {Song}},\ and\ \bibinfo
  {author} {\bibfnamefont {G.}~\bibnamefont {Baym}},\ }\bibfield  {title}
  {\bibinfo {title} {{Phenomenological QCD equation of state for massive
  neutron stars}},\ }\href {https://doi.org/10.1103/PhysRevD.91.045003}
  {\bibfield  {journal} {\bibinfo  {journal} {Phys. Rev. D}\ }\textbf {\bibinfo
  {volume} {91}},\ \bibinfo {pages} {045003} (\bibinfo {year} {2015})},\
  \Eprint {https://arxiv.org/abs/1412.1108} {arXiv:1412.1108 [hep-ph]}
  \BibitemShut {NoStop}%
\bibitem [{\citenamefont {Baym}\ \emph {et~al.}(2018)\citenamefont {Baym},
  \citenamefont {Hatsuda}, \citenamefont {Kojo}, \citenamefont {Powell},
  \citenamefont {Song},\ and\ \citenamefont {Takatsuka}}]{Baym:2017whm}%
  \BibitemOpen
  \bibfield  {author} {\bibinfo {author} {\bibfnamefont {G.}~\bibnamefont
  {Baym}}, \bibinfo {author} {\bibfnamefont {T.}~\bibnamefont {Hatsuda}},
  \bibinfo {author} {\bibfnamefont {T.}~\bibnamefont {Kojo}}, \bibinfo {author}
  {\bibfnamefont {P.~D.}\ \bibnamefont {Powell}}, \bibinfo {author}
  {\bibfnamefont {Y.}~\bibnamefont {Song}},\ and\ \bibinfo {author}
  {\bibfnamefont {T.}~\bibnamefont {Takatsuka}},\ }\bibfield  {title} {\bibinfo
  {title} {{From hadrons to quarks in neutron stars: a review}},\ }\href
  {https://doi.org/10.1088/1361-6633/aaae14} {\bibfield  {journal} {\bibinfo
  {journal} {Rept. Prog. Phys.}\ }\textbf {\bibinfo {volume} {81}},\ \bibinfo
  {pages} {056902} (\bibinfo {year} {2018})},\ \Eprint
  {https://arxiv.org/abs/1707.04966} {arXiv:1707.04966 [astro-ph.HE]}
  \BibitemShut {NoStop}%
\bibitem [{\citenamefont {Baym}\ \emph {et~al.}(2019)\citenamefont {Baym},
  \citenamefont {Furusawa}, \citenamefont {Hatsuda}, \citenamefont {Kojo},\
  and\ \citenamefont {Togashi}}]{Baym:2019iky}%
  \BibitemOpen
  \bibfield  {author} {\bibinfo {author} {\bibfnamefont {G.}~\bibnamefont
  {Baym}}, \bibinfo {author} {\bibfnamefont {S.}~\bibnamefont {Furusawa}},
  \bibinfo {author} {\bibfnamefont {T.}~\bibnamefont {Hatsuda}}, \bibinfo
  {author} {\bibfnamefont {T.}~\bibnamefont {Kojo}},\ and\ \bibinfo {author}
  {\bibfnamefont {H.}~\bibnamefont {Togashi}},\ }\bibfield  {title} {\bibinfo
  {title} {{New Neutron Star Equation of State with Quark-Hadron Crossover}},\
  }\href {https://doi.org/10.3847/1538-4357/ab441e} {\bibfield  {journal}
  {\bibinfo  {journal} {Astrophys. J.}\ }\textbf {\bibinfo {volume} {885}},\
  \bibinfo {pages} {42} (\bibinfo {year} {2019})},\ \Eprint
  {https://arxiv.org/abs/1903.08963} {arXiv:1903.08963 [astro-ph.HE]}
  \BibitemShut {NoStop}%
\bibitem [{\citenamefont {Blaschke}\ \emph {et~al.}(2022)\citenamefont
  {Blaschke}, \citenamefont {Hanu},\ and\ \citenamefont
  {Liebing}}]{Blaschke:2021poc}%
  \BibitemOpen
  \bibfield  {author} {\bibinfo {author} {\bibfnamefont {D.}~\bibnamefont
  {Blaschke}}, \bibinfo {author} {\bibfnamefont {E.~O.}\ \bibnamefont {Hanu}},\
  and\ \bibinfo {author} {\bibfnamefont {S.}~\bibnamefont {Liebing}},\
  }\bibfield  {title} {\bibinfo {title} {{Neutron stars with crossover to color
  superconducting quark matter}},\ }\href
  {https://doi.org/10.1103/PhysRevC.105.035804} {\bibfield  {journal} {\bibinfo
   {journal} {Phys. Rev. C}\ }\textbf {\bibinfo {volume} {105}},\ \bibinfo
  {pages} {035804} (\bibinfo {year} {2022})},\ \Eprint
  {https://arxiv.org/abs/2112.12145} {arXiv:2112.12145 [nucl-th]} \BibitemShut
  {NoStop}%
\bibitem [{\citenamefont {Kojo}\ \emph {et~al.}(2022)\citenamefont {Kojo},
  \citenamefont {Baym},\ and\ \citenamefont {Hatsuda}}]{Kojo:2021wax}%
  \BibitemOpen
  \bibfield  {author} {\bibinfo {author} {\bibfnamefont {T.}~\bibnamefont
  {Kojo}}, \bibinfo {author} {\bibfnamefont {G.}~\bibnamefont {Baym}},\ and\
  \bibinfo {author} {\bibfnamefont {T.}~\bibnamefont {Hatsuda}},\ }\bibfield
  {title} {\bibinfo {title} {{Implications of NICER for Neutron Star Matter:
  The QHC21 Equation of State}},\ }\href
  {https://doi.org/10.3847/1538-4357/ac7876} {\bibfield  {journal} {\bibinfo
  {journal} {Astrophys. J.}\ }\textbf {\bibinfo {volume} {934}},\ \bibinfo
  {pages} {46} (\bibinfo {year} {2022})},\ \Eprint
  {https://arxiv.org/abs/2111.11919} {arXiv:2111.11919 [astro-ph.HE]}
  \BibitemShut {NoStop}%
\bibitem [{\citenamefont {Minamikawa}\ \emph {et~al.}(2021)\citenamefont
  {Minamikawa}, \citenamefont {Kojo},\ and\ \citenamefont
  {Harada}}]{Minamikawa:2020jfj}%
  \BibitemOpen
  \bibfield  {author} {\bibinfo {author} {\bibfnamefont {T.}~\bibnamefont
  {Minamikawa}}, \bibinfo {author} {\bibfnamefont {T.}~\bibnamefont {Kojo}},\
  and\ \bibinfo {author} {\bibfnamefont {M.}~\bibnamefont {Harada}},\
  }\bibfield  {title} {\bibinfo {title} {{Quark-hadron crossover equations of
  state for neutron stars: constraining the chiral invariant mass in a parity
  doublet model}},\ }\href {https://doi.org/10.1103/PhysRevC.103.045205}
  {\bibfield  {journal} {\bibinfo  {journal} {Phys. Rev. C}\ }\textbf {\bibinfo
  {volume} {103}},\ \bibinfo {pages} {045205} (\bibinfo {year} {2021})},\
  \Eprint {https://arxiv.org/abs/2011.13684} {arXiv:2011.13684 [nucl-th]}
  \BibitemShut {NoStop}%
\bibitem [{\citenamefont {Kong}\ \emph {et~al.}(2023)\citenamefont {Kong},
  \citenamefont {Minamikawa},\ and\ \citenamefont {Harada}}]{Kong:2023nue}%
  \BibitemOpen
  \bibfield  {author} {\bibinfo {author} {\bibfnamefont {Y.~K.}\ \bibnamefont
  {Kong}}, \bibinfo {author} {\bibfnamefont {T.}~\bibnamefont {Minamikawa}},\
  and\ \bibinfo {author} {\bibfnamefont {M.}~\bibnamefont {Harada}},\
  }\bibfield  {title} {\bibinfo {title} {{Neutron star matter based on a parity
  doublet model including the a0(980) meson}},\ }\href
  {https://doi.org/10.1103/PhysRevC.108.055206} {\bibfield  {journal} {\bibinfo
   {journal} {Phys. Rev. C}\ }\textbf {\bibinfo {volume} {108}},\ \bibinfo
  {pages} {055206} (\bibinfo {year} {2023})},\ \Eprint
  {https://arxiv.org/abs/2306.08140} {arXiv:2306.08140 [nucl-th]} \BibitemShut
  {NoStop}%
\bibitem [{\citenamefont {Kong}\ \emph {et~al.}(2025)\citenamefont {Kong},
  \citenamefont {Gao},\ and\ \citenamefont {Harada}}]{Kong:2025dwl}%
  \BibitemOpen
  \bibfield  {author} {\bibinfo {author} {\bibfnamefont {Y.-K.}\ \bibnamefont
  {Kong}}, \bibinfo {author} {\bibfnamefont {B.}~\bibnamefont {Gao}},\ and\
  \bibinfo {author} {\bibfnamefont {M.}~\bibnamefont {Harada}},\ }\href@noop {}
  {\bibinfo {title} {{Chiral Invariant Mass Constraints from HESS J1731 347 in
  an Extended Parity Doublet Model with Isovector Scalar Meson}}} (\bibinfo
  {year} {2025}),\ \Eprint {https://arxiv.org/abs/2506.16684} {arXiv:2506.16684
  [nucl-th]} \BibitemShut {NoStop}%
\bibitem [{\citenamefont {McLerran}\ and\ \citenamefont
  {Pisarski}(2007)}]{McLerran:2007qj}%
  \BibitemOpen
  \bibfield  {author} {\bibinfo {author} {\bibfnamefont {L.}~\bibnamefont
  {McLerran}}\ and\ \bibinfo {author} {\bibfnamefont {R.~D.}\ \bibnamefont
  {Pisarski}},\ }\bibfield  {title} {\bibinfo {title} {{Phases of cold, dense
  quarks at large N(c)}},\ }\href
  {https://doi.org/10.1016/j.nuclphysa.2007.08.013} {\bibfield  {journal}
  {\bibinfo  {journal} {Nucl. Phys. A}\ }\textbf {\bibinfo {volume} {796}},\
  \bibinfo {pages} {83} (\bibinfo {year} {2007})},\ \Eprint
  {https://arxiv.org/abs/0706.2191} {arXiv:0706.2191 [hep-ph]} \BibitemShut
  {NoStop}%
\bibitem [{\citenamefont {McLerran}\ \emph {et~al.}(2009)\citenamefont
  {McLerran}, \citenamefont {Redlich},\ and\ \citenamefont
  {Sasaki}}]{McLerran:2008ua}%
  \BibitemOpen
  \bibfield  {author} {\bibinfo {author} {\bibfnamefont {L.}~\bibnamefont
  {McLerran}}, \bibinfo {author} {\bibfnamefont {K.}~\bibnamefont {Redlich}},\
  and\ \bibinfo {author} {\bibfnamefont {C.}~\bibnamefont {Sasaki}},\
  }\bibfield  {title} {\bibinfo {title} {{Quarkyonic Matter and Chiral Symmetry
  Breaking}},\ }\href {https://doi.org/10.1016/j.nuclphysa.2009.04.001}
  {\bibfield  {journal} {\bibinfo  {journal} {Nucl. Phys. A}\ }\textbf
  {\bibinfo {volume} {824}},\ \bibinfo {pages} {86} (\bibinfo {year} {2009})},\
  \Eprint {https://arxiv.org/abs/0812.3585} {arXiv:0812.3585 [hep-ph]}
  \BibitemShut {NoStop}%
\bibitem [{\citenamefont {Hidaka}\ \emph {et~al.}(2008)\citenamefont {Hidaka},
  \citenamefont {McLerran},\ and\ \citenamefont {Pisarski}}]{Hidaka:2008yy}%
  \BibitemOpen
  \bibfield  {author} {\bibinfo {author} {\bibfnamefont {Y.}~\bibnamefont
  {Hidaka}}, \bibinfo {author} {\bibfnamefont {L.~D.}\ \bibnamefont
  {McLerran}},\ and\ \bibinfo {author} {\bibfnamefont {R.~D.}\ \bibnamefont
  {Pisarski}},\ }\bibfield  {title} {\bibinfo {title} {{Baryons and the phase
  diagram for a large number of colors and flavors}},\ }\href
  {https://doi.org/10.1016/j.nuclphysa.2008.05.009} {\bibfield  {journal}
  {\bibinfo  {journal} {Nucl. Phys. A}\ }\textbf {\bibinfo {volume} {808}},\
  \bibinfo {pages} {117} (\bibinfo {year} {2008})},\ \Eprint
  {https://arxiv.org/abs/0803.0279} {arXiv:0803.0279 [hep-ph]} \BibitemShut
  {NoStop}%
\bibitem [{\citenamefont {Fukushima}\ and\ \citenamefont
  {Kojo}(2016)}]{Fukushima:2015bda}%
  \BibitemOpen
  \bibfield  {author} {\bibinfo {author} {\bibfnamefont {K.}~\bibnamefont
  {Fukushima}}\ and\ \bibinfo {author} {\bibfnamefont {T.}~\bibnamefont
  {Kojo}},\ }\bibfield  {title} {\bibinfo {title} {{The Quarkyonic Star}},\
  }\href {https://doi.org/10.3847/0004-637X/817/2/180} {\bibfield  {journal}
  {\bibinfo  {journal} {Astrophys. J.}\ }\textbf {\bibinfo {volume} {817}},\
  \bibinfo {pages} {180} (\bibinfo {year} {2016})},\ \Eprint
  {https://arxiv.org/abs/1509.00356} {arXiv:1509.00356 [nucl-th]} \BibitemShut
  {NoStop}%
\bibitem [{\citenamefont {Duarte}\ \emph {et~al.}(2021)\citenamefont {Duarte},
  \citenamefont {Hernandez-Ortiz}, \citenamefont {Jeong},\ and\ \citenamefont
  {McLerran}}]{Duarte:2021tsx}%
  \BibitemOpen
  \bibfield  {author} {\bibinfo {author} {\bibfnamefont {D.~C.}\ \bibnamefont
  {Duarte}}, \bibinfo {author} {\bibfnamefont {S.}~\bibnamefont
  {Hernandez-Ortiz}}, \bibinfo {author} {\bibfnamefont {K.~S.}\ \bibnamefont
  {Jeong}},\ and\ \bibinfo {author} {\bibfnamefont {L.~D.}\ \bibnamefont
  {McLerran}},\ }\bibfield  {title} {\bibinfo {title} {{Quarkyonic effective
  field theory, quark-nucleon duality, and ghosts}},\ }\href
  {https://doi.org/10.1103/PhysRevD.104.L091901} {\bibfield  {journal}
  {\bibinfo  {journal} {Phys. Rev. D}\ }\textbf {\bibinfo {volume} {104}},\
  \bibinfo {pages} {L091901} (\bibinfo {year} {2021})},\ \Eprint
  {https://arxiv.org/abs/2103.05679} {arXiv:2103.05679 [nucl-th]} \BibitemShut
  {NoStop}%
\bibitem [{\citenamefont {Kojo}(2021)}]{Kojo:2021ugu}%
  \BibitemOpen
  \bibfield  {author} {\bibinfo {author} {\bibfnamefont {T.}~\bibnamefont
  {Kojo}},\ }\bibfield  {title} {\bibinfo {title} {{Stiffening of matter in
  quark-hadron continuity}},\ }\href
  {https://doi.org/10.1103/PhysRevD.104.074005} {\bibfield  {journal} {\bibinfo
   {journal} {Phys. Rev. D}\ }\textbf {\bibinfo {volume} {104}},\ \bibinfo
  {pages} {074005} (\bibinfo {year} {2021})},\ \Eprint
  {https://arxiv.org/abs/2106.06687} {arXiv:2106.06687 [nucl-th]} \BibitemShut
  {NoStop}%
\bibitem [{\citenamefont {McLerran}\ and\ \citenamefont
  {Reddy}(2019)}]{McLerran:2018hbz}%
  \BibitemOpen
  \bibfield  {author} {\bibinfo {author} {\bibfnamefont {L.}~\bibnamefont
  {McLerran}}\ and\ \bibinfo {author} {\bibfnamefont {S.}~\bibnamefont
  {Reddy}},\ }\bibfield  {title} {\bibinfo {title} {{Quarkyonic Matter and
  Neutron Stars}},\ }\href {https://doi.org/10.1103/PhysRevLett.122.122701}
  {\bibfield  {journal} {\bibinfo  {journal} {Phys. Rev. Lett.}\ }\textbf
  {\bibinfo {volume} {122}},\ \bibinfo {pages} {122701} (\bibinfo {year}
  {2019})},\ \Eprint {https://arxiv.org/abs/1811.12503} {arXiv:1811.12503
  [nucl-th]} \BibitemShut {NoStop}%
\bibitem [{\citenamefont {Jeong}\ \emph {et~al.}(2020)\citenamefont {Jeong},
  \citenamefont {McLerran},\ and\ \citenamefont {Sen}}]{Jeong:2019lhv}%
  \BibitemOpen
  \bibfield  {author} {\bibinfo {author} {\bibfnamefont {K.~S.}\ \bibnamefont
  {Jeong}}, \bibinfo {author} {\bibfnamefont {L.}~\bibnamefont {McLerran}},\
  and\ \bibinfo {author} {\bibfnamefont {S.}~\bibnamefont {Sen}},\ }\bibfield
  {title} {\bibinfo {title} {{Dynamically generated momentum space shell
  structure of quarkyonic matter via an excluded volume model}},\ }\href
  {https://doi.org/10.1103/PhysRevC.101.035201} {\bibfield  {journal} {\bibinfo
   {journal} {Phys. Rev. C}\ }\textbf {\bibinfo {volume} {101}},\ \bibinfo
  {pages} {035201} (\bibinfo {year} {2020})},\ \Eprint
  {https://arxiv.org/abs/1908.04799} {arXiv:1908.04799 [nucl-th]} \BibitemShut
  {NoStop}%
\bibitem [{\citenamefont {Sen}\ and\ \citenamefont
  {Warrington}(2021)}]{Sen:2020peq}%
  \BibitemOpen
  \bibfield  {author} {\bibinfo {author} {\bibfnamefont {S.}~\bibnamefont
  {Sen}}\ and\ \bibinfo {author} {\bibfnamefont {N.~C.}\ \bibnamefont
  {Warrington}},\ }\bibfield  {title} {\bibinfo {title} {{Finite-temperature
  quarkyonic matter with an excluded volume model}},\ }\href
  {https://doi.org/10.1016/j.nuclphysa.2020.122059} {\bibfield  {journal}
  {\bibinfo  {journal} {Nucl. Phys. A}\ }\textbf {\bibinfo {volume} {1006}},\
  \bibinfo {pages} {122059} (\bibinfo {year} {2021})},\ \Eprint
  {https://arxiv.org/abs/2002.11133} {arXiv:2002.11133 [nucl-th]} \BibitemShut
  {NoStop}%
\bibitem [{\citenamefont {Fujimoto}\ \emph
  {et~al.}(2024{\natexlab{a}})\citenamefont {Fujimoto}, \citenamefont {Kojo},\
  and\ \citenamefont {McLerran}}]{Fujimoto:2023mzy}%
  \BibitemOpen
  \bibfield  {author} {\bibinfo {author} {\bibfnamefont {Y.}~\bibnamefont
  {Fujimoto}}, \bibinfo {author} {\bibfnamefont {T.}~\bibnamefont {Kojo}},\
  and\ \bibinfo {author} {\bibfnamefont {L.~D.}\ \bibnamefont {McLerran}},\
  }\bibfield  {title} {\bibinfo {title} {{Momentum Shell in Quarkyonic Matter
  from Explicit Duality: A Dual Model for Cold, Dense QCD}},\ }\href
  {https://doi.org/10.1103/PhysRevLett.132.112701} {\bibfield  {journal}
  {\bibinfo  {journal} {Phys. Rev. Lett.}\ }\textbf {\bibinfo {volume} {132}},\
  \bibinfo {pages} {112701} (\bibinfo {year} {2024}{\natexlab{a}})},\ \Eprint
  {https://arxiv.org/abs/2306.04304} {arXiv:2306.04304 [nucl-th]} \BibitemShut
  {NoStop}%
\bibitem [{\citenamefont {Fujimoto}\ \emph
  {et~al.}(2024{\natexlab{b}})\citenamefont {Fujimoto}, \citenamefont {Kojo},\
  and\ \citenamefont {McLerran}}]{Fujimoto:2024doc}%
  \BibitemOpen
  \bibfield  {author} {\bibinfo {author} {\bibfnamefont {Y.}~\bibnamefont
  {Fujimoto}}, \bibinfo {author} {\bibfnamefont {T.}~\bibnamefont {Kojo}},\
  and\ \bibinfo {author} {\bibfnamefont {L.}~\bibnamefont {McLerran}},\
  }\href@noop {} {\bibinfo {title} {{Quarkyonic matter pieces together the
  hyperon puzzle}}} (\bibinfo {year} {2024}{\natexlab{b}}),\ \Eprint
  {https://arxiv.org/abs/2410.22758} {arXiv:2410.22758 [nucl-th]} \BibitemShut
  {NoStop}%
\bibitem [{\citenamefont {Kojo}(2024)}]{Kojo:2024ejq}%
  \BibitemOpen
  \bibfield  {author} {\bibinfo {author} {\bibfnamefont {T.}~\bibnamefont
  {Kojo}},\ }\href@noop {} {\bibinfo {title} {{Stiffening of matter in
  quark-hadron continuity: a mini-review}}} (\bibinfo {year} {2024}),\ \Eprint
  {https://arxiv.org/abs/2412.20442} {arXiv:2412.20442 [nucl-th]} \BibitemShut
  {NoStop}%
\bibitem [{\citenamefont {Ivanytskyi}(2025)}]{Ivanytskyi:2025cnn}%
  \BibitemOpen
  \bibfield  {author} {\bibinfo {author} {\bibfnamefont {O.}~\bibnamefont
  {Ivanytskyi}},\ }\href@noop {} {\bibinfo {title} {{Quarkyonic picture of
  isospin QCD}}} (\bibinfo {year} {2025}),\ \Eprint
  {https://arxiv.org/abs/2505.07076} {arXiv:2505.07076 [nucl-th]} \BibitemShut
  {NoStop}%
\bibitem [{\citenamefont {Gao}\ and\ \citenamefont
  {Yoshida}(2025)}]{Gao:2025rbq}%
  \BibitemOpen
  \bibfield  {author} {\bibinfo {author} {\bibfnamefont {B.}~\bibnamefont
  {Gao}}\ and\ \bibinfo {author} {\bibfnamefont {K.}~\bibnamefont {Yoshida}},\
  }\href@noop {} {\bibinfo {title} {{Ferromagnetic instabilities in quarkyonic
  matter}}} (\bibinfo {year} {2025}),\ \Eprint
  {https://arxiv.org/abs/2507.06577} {arXiv:2507.06577 [nucl-th]} \BibitemShut
  {NoStop}%
\bibitem [{\citenamefont {Kojo}(2025)}]{Kojo:2025vcq}%
  \BibitemOpen
  \bibfield  {author} {\bibinfo {author} {\bibfnamefont {T.}~\bibnamefont
  {Kojo}},\ }\bibfield  {title} {\bibinfo {title} {{Stiffening of matter in
  quark{\textendash}hadron continuity: A mini-review}},\ }\href
  {https://doi.org/10.1016/j.jspc.2025.100088} {\bibfield  {journal} {\bibinfo
  {journal} {J. Subatomic Part. Cosmol.}\ }\textbf {\bibinfo {volume} {4}},\
  \bibinfo {pages} {100088} (\bibinfo {year} {2025})},\ \Eprint
  {https://arxiv.org/abs/2412.20442} {arXiv:2412.20442 [nucl-th]} \BibitemShut
  {NoStop}%
\bibitem [{\citenamefont {Fujimoto}\ \emph {et~al.}(2025)\citenamefont
  {Fujimoto}, \citenamefont {Kojo},\ and\ \citenamefont
  {McLerran}}]{Fujimoto:2025trx}%
  \BibitemOpen
  \bibfield  {author} {\bibinfo {author} {\bibfnamefont {Y.}~\bibnamefont
  {Fujimoto}}, \bibinfo {author} {\bibfnamefont {T.}~\bibnamefont {Kojo}},\
  and\ \bibinfo {author} {\bibfnamefont {L.}~\bibnamefont {McLerran}},\
  }\bibfield  {title} {\bibinfo {title} {{Quarkyonic solution to the hyperon
  puzzle}},\ }\href {https://doi.org/10.1051/epjconf/202531607007} {\bibfield
  {journal} {\bibinfo  {journal} {EPJ Web Conf.}\ }\textbf {\bibinfo {volume}
  {316}},\ \bibinfo {pages} {07007} (\bibinfo {year} {2025})}\BibitemShut
  {NoStop}%
\bibitem [{\citenamefont {Koch}\ and\ \citenamefont
  {Vovchenko}(2023)}]{Koch:2022act}%
  \BibitemOpen
  \bibfield  {author} {\bibinfo {author} {\bibfnamefont {V.}~\bibnamefont
  {Koch}}\ and\ \bibinfo {author} {\bibfnamefont {V.}~\bibnamefont
  {Vovchenko}},\ }\bibfield  {title} {\bibinfo {title} {{Quarkyonic or
  baryquark matter? On the dynamical generation of momentum space shell
  structure}},\ }\href {https://doi.org/10.1016/j.physletb.2023.137942}
  {\bibfield  {journal} {\bibinfo  {journal} {Phys. Lett. B}\ }\textbf
  {\bibinfo {volume} {841}},\ \bibinfo {pages} {137942} (\bibinfo {year}
  {2023})},\ \Eprint {https://arxiv.org/abs/2211.14674} {arXiv:2211.14674
  [nucl-th]} \BibitemShut {NoStop}%
\bibitem [{\citenamefont {Poberezhnyuk}\ \emph {et~al.}(2023)\citenamefont
  {Poberezhnyuk}, \citenamefont {Stoecker},\ and\ \citenamefont
  {Vovchenko}}]{Poberezhnyuk:2023rct}%
  \BibitemOpen
  \bibfield  {author} {\bibinfo {author} {\bibfnamefont {R.~V.}\ \bibnamefont
  {Poberezhnyuk}}, \bibinfo {author} {\bibfnamefont {H.}~\bibnamefont
  {Stoecker}},\ and\ \bibinfo {author} {\bibfnamefont {V.}~\bibnamefont
  {Vovchenko}},\ }\bibfield  {title} {\bibinfo {title} {{Quarkyonic matter with
  quantum van der Waals theory}},\ }\href
  {https://doi.org/10.1103/PhysRevC.108.045202} {\bibfield  {journal} {\bibinfo
   {journal} {Phys. Rev. C}\ }\textbf {\bibinfo {volume} {108}},\ \bibinfo
  {pages} {045202} (\bibinfo {year} {2023})},\ \Eprint
  {https://arxiv.org/abs/2307.13532} {arXiv:2307.13532 [nucl-th]} \BibitemShut
  {NoStop}%
\bibitem [{\citenamefont {Koch}\ and\ \citenamefont
  {Vovchenko}(2025)}]{Koch:2025sfv}%
  \BibitemOpen
  \bibfield  {author} {\bibinfo {author} {\bibfnamefont {V.}~\bibnamefont
  {Koch}}\ and\ \bibinfo {author} {\bibfnamefont {V.}~\bibnamefont
  {Vovchenko}},\ }\bibfield  {title} {\bibinfo {title} {{Quarkyonic or
  baryquark matter}},\ }\href {https://doi.org/10.1016/j.jspc.2025.100025}
  {\bibfield  {journal} {\bibinfo  {journal} {J. Subatomic Part. Cosmol.}\
  }\textbf {\bibinfo {volume} {3}},\ \bibinfo {pages} {100025} (\bibinfo {year}
  {2025})}\BibitemShut {NoStop}%
\bibitem [{\citenamefont {Detar}\ and\ \citenamefont
  {Kunihiro}(1989)}]{Detar:1988kn}%
  \BibitemOpen
  \bibfield  {author} {\bibinfo {author} {\bibfnamefont {C.~E.}\ \bibnamefont
  {Detar}}\ and\ \bibinfo {author} {\bibfnamefont {T.}~\bibnamefont
  {Kunihiro}},\ }\bibfield  {title} {\bibinfo {title} {{Linear $\sigma$ Model
  With Parity Doubling}},\ }\href {https://doi.org/10.1103/PhysRevD.39.2805}
  {\bibfield  {journal} {\bibinfo  {journal} {Phys. Rev. D}\ }\textbf {\bibinfo
  {volume} {39}},\ \bibinfo {pages} {2805} (\bibinfo {year}
  {1989})}\BibitemShut {NoStop}%
\bibitem [{\citenamefont {Jido}\ \emph {et~al.}(2001)\citenamefont {Jido},
  \citenamefont {Oka},\ and\ \citenamefont {Hosaka}}]{Jido:2001nt}%
  \BibitemOpen
  \bibfield  {author} {\bibinfo {author} {\bibfnamefont {D.}~\bibnamefont
  {Jido}}, \bibinfo {author} {\bibfnamefont {M.}~\bibnamefont {Oka}},\ and\
  \bibinfo {author} {\bibfnamefont {A.}~\bibnamefont {Hosaka}},\ }\bibfield
  {title} {\bibinfo {title} {{Chiral symmetry of baryons}},\ }\href
  {https://doi.org/10.1143/PTP.106.873} {\bibfield  {journal} {\bibinfo
  {journal} {Prog. Theor. Phys.}\ }\textbf {\bibinfo {volume} {106}},\ \bibinfo
  {pages} {873} (\bibinfo {year} {2001})},\ \Eprint
  {https://arxiv.org/abs/hep-ph/0110005} {arXiv:hep-ph/0110005} \BibitemShut
  {NoStop}%
\bibitem [{\citenamefont {Marczenko}\ \emph {et~al.}(2019)\citenamefont
  {Marczenko}, \citenamefont {Blaschke}, \citenamefont {Redlich},\ and\
  \citenamefont {Sasaki}}]{universe5080180}%
  \BibitemOpen
  \bibfield  {author} {\bibinfo {author} {\bibfnamefont {M.}~\bibnamefont
  {Marczenko}}, \bibinfo {author} {\bibfnamefont {D.}~\bibnamefont {Blaschke}},
  \bibinfo {author} {\bibfnamefont {K.}~\bibnamefont {Redlich}},\ and\ \bibinfo
  {author} {\bibfnamefont {C.}~\bibnamefont {Sasaki}},\ }\bibfield  {title}
  {\bibinfo {title} {Parity doubling and the dense-matter phase diagram under
  constraints from multi-messenger astronomy},\ }\bibfield  {journal} {\bibinfo
   {journal} {Universe}\ }\textbf {\bibinfo {volume} {5}},\ \href
  {https://doi.org/10.3390/universe5080180} {10.3390/universe5080180} (\bibinfo
  {year} {2019})\BibitemShut {NoStop}%
\bibitem [{\citenamefont {Yamazaki}\ and\ \citenamefont
  {Harada}(2019)}]{PhysRevC.100.025205}%
  \BibitemOpen
  \bibfield  {author} {\bibinfo {author} {\bibfnamefont {T.}~\bibnamefont
  {Yamazaki}}\ and\ \bibinfo {author} {\bibfnamefont {M.}~\bibnamefont
  {Harada}},\ }\bibfield  {title} {\bibinfo {title} {Constraint to chiral
  invariant masses of nucleons from gw170817 in an extended parity doublet
  model},\ }\href {https://doi.org/10.1103/PhysRevC.100.025205} {\bibfield
  {journal} {\bibinfo  {journal} {Phys. Rev. C}\ }\textbf {\bibinfo {volume}
  {100}},\ \bibinfo {pages} {025205} (\bibinfo {year} {2019})}\BibitemShut
  {NoStop}%
\bibitem [{\citenamefont {Mukherjee}\ \emph {et~al.}(2017)\citenamefont
  {Mukherjee}, \citenamefont {Schramm}, \citenamefont {Steinheimer},\ and\
  \citenamefont {Dexheimer}}]{Mukherjee:2017jzi}%
  \BibitemOpen
  \bibfield  {author} {\bibinfo {author} {\bibfnamefont {A.}~\bibnamefont
  {Mukherjee}}, \bibinfo {author} {\bibfnamefont {S.}~\bibnamefont {Schramm}},
  \bibinfo {author} {\bibfnamefont {J.}~\bibnamefont {Steinheimer}},\ and\
  \bibinfo {author} {\bibfnamefont {V.}~\bibnamefont {Dexheimer}},\ }\bibfield
  {title} {\bibinfo {title} {{The application of the Quark-Hadron Chiral
  Parity-Doublet Model to neutron star matter}},\ }\href
  {https://doi.org/10.1051/0004-6361/201731505} {\bibfield  {journal} {\bibinfo
   {journal} {Astron. Astrophys.}\ }\textbf {\bibinfo {volume} {608}},\
  \bibinfo {pages} {A110} (\bibinfo {year} {2017})},\ \Eprint
  {https://arxiv.org/abs/1706.09191} {arXiv:1706.09191 [nucl-th]} \BibitemShut
  {NoStop}%
\bibitem [{\citenamefont {Marczenko}\ \emph
  {et~al.}(2022{\natexlab{a}})\citenamefont {Marczenko}, \citenamefont
  {Redlich},\ and\ \citenamefont {Sasaki}}]{Marczenko:2021uaj}%
  \BibitemOpen
  \bibfield  {author} {\bibinfo {author} {\bibfnamefont {M.}~\bibnamefont
  {Marczenko}}, \bibinfo {author} {\bibfnamefont {K.}~\bibnamefont {Redlich}},\
  and\ \bibinfo {author} {\bibfnamefont {C.}~\bibnamefont {Sasaki}},\
  }\bibfield  {title} {\bibinfo {title} {{Reconciling Multi-messenger
  Constraints with Chiral Symmetry Restoration}},\ }\href
  {https://doi.org/10.3847/2041-8213/ac4b61} {\bibfield  {journal} {\bibinfo
  {journal} {Astrophys. J. Lett.}\ }\textbf {\bibinfo {volume} {925}},\
  \bibinfo {pages} {L23} (\bibinfo {year} {2022}{\natexlab{a}})},\ \Eprint
  {https://arxiv.org/abs/2110.11056} {arXiv:2110.11056 [nucl-th]} \BibitemShut
  {NoStop}%
\bibitem [{\citenamefont {Marczenko}\ \emph
  {et~al.}(2022{\natexlab{b}})\citenamefont {Marczenko}, \citenamefont
  {Redlich},\ and\ \citenamefont {Sasaki}}]{Marczenko:2022hyt}%
  \BibitemOpen
  \bibfield  {author} {\bibinfo {author} {\bibfnamefont {M.}~\bibnamefont
  {Marczenko}}, \bibinfo {author} {\bibfnamefont {K.}~\bibnamefont {Redlich}},\
  and\ \bibinfo {author} {\bibfnamefont {C.}~\bibnamefont {Sasaki}},\
  }\bibfield  {title} {\bibinfo {title} {{Chiral symmetry restoration and
  \ensuremath{\Delta} matter formation in neutron stars}},\ }\href
  {https://doi.org/10.1103/PhysRevD.105.103009} {\bibfield  {journal} {\bibinfo
   {journal} {Phys. Rev. D}\ }\textbf {\bibinfo {volume} {105}},\ \bibinfo
  {pages} {103009} (\bibinfo {year} {2022}{\natexlab{b}})},\ \Eprint
  {https://arxiv.org/abs/2203.00269} {arXiv:2203.00269 [nucl-th]} \BibitemShut
  {NoStop}%
\bibitem [{\citenamefont {Gao}\ \emph {et~al.}(2024{\natexlab{b}})\citenamefont
  {Gao}, \citenamefont {Yan},\ and\ \citenamefont {Harada}}]{Gao:2024chh}%
  \BibitemOpen
  \bibfield  {author} {\bibinfo {author} {\bibfnamefont {B.}~\bibnamefont
  {Gao}}, \bibinfo {author} {\bibfnamefont {Y.}~\bibnamefont {Yan}},\ and\
  \bibinfo {author} {\bibfnamefont {M.}~\bibnamefont {Harada}},\ }\bibfield
  {title} {\bibinfo {title} {{Reconciling constraints from the supernova
  remnant HESS J1731-347 with the parity doublet model}},\ }\href
  {https://doi.org/10.1103/PhysRevC.109.065807} {\bibfield  {journal} {\bibinfo
   {journal} {Phys. Rev. C}\ }\textbf {\bibinfo {volume} {109}},\ \bibinfo
  {pages} {065807} (\bibinfo {year} {2024}{\natexlab{b}})},\ \Eprint
  {https://arxiv.org/abs/2404.04786} {arXiv:2404.04786 [nucl-th]} \BibitemShut
  {NoStop}%
\bibitem [{\citenamefont {Marczenko}\ \emph {et~al.}(2020)\citenamefont
  {Marczenko}, \citenamefont {Blaschke}, \citenamefont {Redlich},\ and\
  \citenamefont {Sasaki}}]{Marczenko:2020jma}%
  \BibitemOpen
  \bibfield  {author} {\bibinfo {author} {\bibfnamefont {M.}~\bibnamefont
  {Marczenko}}, \bibinfo {author} {\bibfnamefont {D.}~\bibnamefont {Blaschke}},
  \bibinfo {author} {\bibfnamefont {K.}~\bibnamefont {Redlich}},\ and\ \bibinfo
  {author} {\bibfnamefont {C.}~\bibnamefont {Sasaki}},\ }\bibfield  {title}
  {\bibinfo {title} {{Toward a unified equation of state for multi-messenger
  astronomy}},\ }\href {https://doi.org/10.1051/0004-6361/202038211} {\bibfield
   {journal} {\bibinfo  {journal} {Astron. Astrophys.}\ }\textbf {\bibinfo
  {volume} {643}},\ \bibinfo {pages} {A82} (\bibinfo {year} {2020})},\ \Eprint
  {https://arxiv.org/abs/2004.09566} {arXiv:2004.09566 [astro-ph.HE]}
  \BibitemShut {NoStop}%
\bibitem [{\citenamefont {Eser}\ and\ \citenamefont
  {Blaizot}(2024{\natexlab{a}})}]{Eser:2023oii}%
  \BibitemOpen
  \bibfield  {author} {\bibinfo {author} {\bibfnamefont {J.}~\bibnamefont
  {Eser}}\ and\ \bibinfo {author} {\bibfnamefont {J.-P.}\ \bibnamefont
  {Blaizot}},\ }\bibfield  {title} {\bibinfo {title} {{Thermodynamics of the
  parity-doublet model: Symmetric nuclear matter and the chiral transition}},\
  }\href {https://doi.org/10.1103/PhysRevC.109.045201} {\bibfield  {journal}
  {\bibinfo  {journal} {Phys. Rev. C}\ }\textbf {\bibinfo {volume} {109}},\
  \bibinfo {pages} {045201} (\bibinfo {year} {2024}{\natexlab{a}})},\ \Eprint
  {https://arxiv.org/abs/2309.06566} {arXiv:2309.06566 [nucl-th]} \BibitemShut
  {NoStop}%
\bibitem [{\citenamefont {Eser}\ and\ \citenamefont
  {Blaizot}(2024{\natexlab{b}})}]{Eser:2024xil}%
  \BibitemOpen
  \bibfield  {author} {\bibinfo {author} {\bibfnamefont {J.}~\bibnamefont
  {Eser}}\ and\ \bibinfo {author} {\bibfnamefont {J.-P.}\ \bibnamefont
  {Blaizot}},\ }\bibfield  {title} {\bibinfo {title} {{Thermodynamics of the
  parity-doublet model. II. Asymmetric and neutron matter}},\ }\href
  {https://doi.org/10.1103/PhysRevC.110.065205} {\bibfield  {journal} {\bibinfo
   {journal} {Phys. Rev. C}\ }\textbf {\bibinfo {volume} {110}},\ \bibinfo
  {pages} {065205} (\bibinfo {year} {2024}{\natexlab{b}})},\ \Eprint
  {https://arxiv.org/abs/2408.01302} {arXiv:2408.01302 [nucl-th]} \BibitemShut
  {NoStop}%
\bibitem [{\citenamefont {Shao}\ and\ \citenamefont {Ma}(2022)}]{Shao:2022njr}%
  \BibitemOpen
  \bibfield  {author} {\bibinfo {author} {\bibfnamefont {L.-Q.}\ \bibnamefont
  {Shao}}\ and\ \bibinfo {author} {\bibfnamefont {Y.-L.}\ \bibnamefont {Ma}},\
  }\bibfield  {title} {\bibinfo {title} {{Scale symmetry and composition of
  compact star matter}},\ }\href {https://doi.org/10.1103/PhysRevD.106.014014}
  {\bibfield  {journal} {\bibinfo  {journal} {Phys. Rev. D}\ }\textbf {\bibinfo
  {volume} {106}},\ \bibinfo {pages} {014014} (\bibinfo {year} {2022})},\
  \Eprint {https://arxiv.org/abs/2202.09957} {arXiv:2202.09957 [nucl-th]}
  \BibitemShut {NoStop}%
\bibitem [{\citenamefont {Ma}\ and\ \citenamefont {Ma}(2024)}]{Ma:2023eoz}%
  \BibitemOpen
  \bibfield  {author} {\bibinfo {author} {\bibfnamefont {Y.}~\bibnamefont
  {Ma}}\ and\ \bibinfo {author} {\bibfnamefont {Y.-L.}\ \bibnamefont {Ma}},\
  }\bibfield  {title} {\bibinfo {title} {{Quark structure of isoscalar- and
  isovector-scalar mesons and nuclear matter property}},\ }\href
  {https://doi.org/10.1103/PhysRevD.109.074022} {\bibfield  {journal} {\bibinfo
   {journal} {Phys. Rev. D}\ }\textbf {\bibinfo {volume} {109}},\ \bibinfo
  {pages} {074022} (\bibinfo {year} {2024})},\ \Eprint
  {https://arxiv.org/abs/2311.07899} {arXiv:2311.07899 [nucl-th]} \BibitemShut
  {NoStop}%
\bibitem [{\citenamefont {Guo}\ \emph {et~al.}(2024)\citenamefont {Guo},
  \citenamefont {Xiong}, \citenamefont {Ma},\ and\ \citenamefont
  {Ma}}]{Guo:2023mhf}%
  \BibitemOpen
  \bibfield  {author} {\bibinfo {author} {\bibfnamefont {L.-J.}\ \bibnamefont
  {Guo}}, \bibinfo {author} {\bibfnamefont {J.-Y.}\ \bibnamefont {Xiong}},
  \bibinfo {author} {\bibfnamefont {Y.}~\bibnamefont {Ma}},\ and\ \bibinfo
  {author} {\bibfnamefont {Y.-L.}\ \bibnamefont {Ma}},\ }\bibfield  {title}
  {\bibinfo {title} {{Insights into Neutron Star Equation of State by Machine
  Learning}},\ }\href {https://doi.org/10.3847/1538-4357/ad2e8d} {\bibfield
  {journal} {\bibinfo  {journal} {Astrophys. J.}\ }\textbf {\bibinfo {volume}
  {965}},\ \bibinfo {pages} {47} (\bibinfo {year} {2024})},\ \Eprint
  {https://arxiv.org/abs/2309.11227} {arXiv:2309.11227 [nucl-th]} \BibitemShut
  {NoStop}%
\bibitem [{\citenamefont {Guo}\ \emph {et~al.}(2025)\citenamefont {Guo},
  \citenamefont {Yang}, \citenamefont {Ma},\ and\ \citenamefont
  {Wu}}]{Guo:2023som}%
  \BibitemOpen
  \bibfield  {author} {\bibinfo {author} {\bibfnamefont {L.-J.}\ \bibnamefont
  {Guo}}, \bibinfo {author} {\bibfnamefont {W.-C.}\ \bibnamefont {Yang}},
  \bibinfo {author} {\bibfnamefont {Y.-L.}\ \bibnamefont {Ma}},\ and\ \bibinfo
  {author} {\bibfnamefont {Y.-L.}\ \bibnamefont {Wu}},\ }\bibfield  {title}
  {\bibinfo {title} {{Probing Hadron-quark Transition Through Binary Neutron
  Star Merger}},\ }\href {https://doi.org/10.1088/1674-4527/adbc37} {\bibfield
  {journal} {\bibinfo  {journal} {Res. Astron. Astrophys.}\ }\textbf {\bibinfo
  {volume} {25}},\ \bibinfo {pages} {035017} (\bibinfo {year} {2025})},\
  \Eprint {https://arxiv.org/abs/2308.01770} {arXiv:2308.01770 [astro-ph.HE]}
  \BibitemShut {NoStop}%
\bibitem [{\citenamefont {Ma}\ and\ \citenamefont {Yang}(2023)}]{Ma:2023ugl}%
  \BibitemOpen
  \bibfield  {author} {\bibinfo {author} {\bibfnamefont {Y.-L.}\ \bibnamefont
  {Ma}}\ and\ \bibinfo {author} {\bibfnamefont {W.-C.}\ \bibnamefont {Yang}},\
  }\bibfield  {title} {\bibinfo {title} {{Topology and Emergent Symmetries in
  Dense Compact Star Matter}},\ }\href {https://doi.org/10.3390/sym15030776}
  {\bibfield  {journal} {\bibinfo  {journal} {Symmetry}\ }\textbf {\bibinfo
  {volume} {15}},\ \bibinfo {pages} {776} (\bibinfo {year} {2023})},\ \Eprint
  {https://arxiv.org/abs/2301.02105} {arXiv:2301.02105 [nucl-th]} \BibitemShut
  {NoStop}%
\bibitem [{\citenamefont {Gao}\ \emph {et~al.}(2024{\natexlab{c}})\citenamefont
  {Gao}, \citenamefont {Kojo},\ and\ \citenamefont {Harada}}]{Gao:2024mew}%
  \BibitemOpen
  \bibfield  {author} {\bibinfo {author} {\bibfnamefont {B.}~\bibnamefont
  {Gao}}, \bibinfo {author} {\bibfnamefont {T.}~\bibnamefont {Kojo}},\ and\
  \bibinfo {author} {\bibfnamefont {M.}~\bibnamefont {Harada}},\ }\bibfield
  {title} {\bibinfo {title} {{Parity doublet model for baryon octets: Ground
  states saturated by good diquarks and the role of bad diquarks for excited
  states}},\ }\href {https://doi.org/10.1103/PhysRevD.110.016016} {\bibfield
  {journal} {\bibinfo  {journal} {Phys. Rev. D}\ }\textbf {\bibinfo {volume}
  {110}},\ \bibinfo {pages} {016016} (\bibinfo {year} {2024}{\natexlab{c}})},\
  \Eprint {https://arxiv.org/abs/2403.18214} {arXiv:2403.18214 [hep-ph]}
  \BibitemShut {NoStop}%
\bibitem [{\citenamefont {Marczenko}\ \emph {et~al.}(2023)\citenamefont
  {Marczenko}, \citenamefont {Redlich},\ and\ \citenamefont
  {Sasaki}}]{Marczenko:2023ohi}%
  \BibitemOpen
  \bibfield  {author} {\bibinfo {author} {\bibfnamefont {M.}~\bibnamefont
  {Marczenko}}, \bibinfo {author} {\bibfnamefont {K.}~\bibnamefont {Redlich}},\
  and\ \bibinfo {author} {\bibfnamefont {C.}~\bibnamefont {Sasaki}},\
  }\bibfield  {title} {\bibinfo {title} {{Fluctuations near the liquid-gas and
  chiral phase transitions in hadronic matter}},\ }\href
  {https://doi.org/10.1103/PhysRevD.107.054046} {\bibfield  {journal} {\bibinfo
   {journal} {Phys. Rev. D}\ }\textbf {\bibinfo {volume} {107}},\ \bibinfo
  {pages} {054046} (\bibinfo {year} {2023})},\ \Eprint
  {https://arxiv.org/abs/2301.09866} {arXiv:2301.09866 [nucl-th]} \BibitemShut
  {NoStop}%
\bibitem [{\citenamefont {Koch}\ \emph {et~al.}(2024)\citenamefont {Koch},
  \citenamefont {Marczenko}, \citenamefont {Redlich},\ and\ \citenamefont
  {Sasaki}}]{Koch:2023oez}%
  \BibitemOpen
  \bibfield  {author} {\bibinfo {author} {\bibfnamefont {V.}~\bibnamefont
  {Koch}}, \bibinfo {author} {\bibfnamefont {M.}~\bibnamefont {Marczenko}},
  \bibinfo {author} {\bibfnamefont {K.}~\bibnamefont {Redlich}},\ and\ \bibinfo
  {author} {\bibfnamefont {C.}~\bibnamefont {Sasaki}},\ }\bibfield  {title}
  {\bibinfo {title} {{Fluctuations and correlations of baryonic chiral
  partners}},\ }\href {https://doi.org/10.1103/PhysRevD.109.014033} {\bibfield
  {journal} {\bibinfo  {journal} {Phys. Rev. D}\ }\textbf {\bibinfo {volume}
  {109}},\ \bibinfo {pages} {014033} (\bibinfo {year} {2024})},\ \Eprint
  {https://arxiv.org/abs/2308.15794} {arXiv:2308.15794 [hep-ph]} \BibitemShut
  {NoStop}%
\bibitem [{\citenamefont {Marczenko}\ \emph {et~al.}(2025)\citenamefont
  {Marczenko}, \citenamefont {Redlich},\ and\ \citenamefont
  {Sasaki}}]{Marczenko:2024nge}%
  \BibitemOpen
  \bibfield  {author} {\bibinfo {author} {\bibfnamefont {M.}~\bibnamefont
  {Marczenko}}, \bibinfo {author} {\bibfnamefont {K.}~\bibnamefont {Redlich}},\
  and\ \bibinfo {author} {\bibfnamefont {C.}~\bibnamefont {Sasaki}},\
  }\bibfield  {title} {\bibinfo {title} {{Probing the nuclear liquid-gas phase
  transition with isospin correlations}},\ }\href
  {https://doi.org/10.1103/4ccc-j338} {\bibfield  {journal} {\bibinfo
  {journal} {Phys. Rev. C}\ }\textbf {\bibinfo {volume} {111}},\ \bibinfo
  {pages} {065203} (\bibinfo {year} {2025})},\ \Eprint
  {https://arxiv.org/abs/2410.21746} {arXiv:2410.21746 [nucl-th]} \BibitemShut
  {NoStop}%
\bibitem [{\citenamefont {Gao}\ and\ \citenamefont
  {Harada}(2025)}]{Gao:2024jlp}%
  \BibitemOpen
  \bibfield  {author} {\bibinfo {author} {\bibfnamefont {B.}~\bibnamefont
  {Gao}}\ and\ \bibinfo {author} {\bibfnamefont {M.}~\bibnamefont {Harada}},\
  }\bibfield  {title} {\bibinfo {title} {{Quarkyonic matter with chiral
  symmetry restoration}},\ }\href {https://doi.org/10.1103/PhysRevD.111.016024}
  {\bibfield  {journal} {\bibinfo  {journal} {Phys. Rev. D}\ }\textbf {\bibinfo
  {volume} {111}},\ \bibinfo {pages} {016024} (\bibinfo {year} {2025})},\
  \Eprint {https://arxiv.org/abs/2410.16649} {arXiv:2410.16649 [nucl-th]}
  \BibitemShut {NoStop}%
\bibitem [{\citenamefont {Motohiro}\ \emph {et~al.}(2015)\citenamefont
  {Motohiro}, \citenamefont {Kim},\ and\ \citenamefont {Harada}}]{Motohiro}%
  \BibitemOpen
  \bibfield  {author} {\bibinfo {author} {\bibfnamefont {Y.}~\bibnamefont
  {Motohiro}}, \bibinfo {author} {\bibfnamefont {Y.}~\bibnamefont {Kim}},\ and\
  \bibinfo {author} {\bibfnamefont {M.}~\bibnamefont {Harada}},\ }\bibfield
  {title} {\bibinfo {title} {Asymmetric nuclear matter in a parity doublet
  model with hidden local symmetry},\ }\href
  {https://doi.org/10.1103/PhysRevC.92.025201} {\bibfield  {journal} {\bibinfo
  {journal} {Phys. Rev. C}\ }\textbf {\bibinfo {volume} {92}},\ \bibinfo
  {pages} {025201} (\bibinfo {year} {2015})}\BibitemShut {NoStop}%
\bibitem [{\citenamefont {Zhao}\ and\ \citenamefont
  {Lattimer}(2020)}]{Zhao:2020dvu}%
  \BibitemOpen
  \bibfield  {author} {\bibinfo {author} {\bibfnamefont {T.}~\bibnamefont
  {Zhao}}\ and\ \bibinfo {author} {\bibfnamefont {J.~M.}\ \bibnamefont
  {Lattimer}},\ }\bibfield  {title} {\bibinfo {title} {{Quarkyonic Matter
  Equation of State in Beta-Equilibrium}},\ }\href
  {https://doi.org/10.1103/PhysRevD.102.023021} {\bibfield  {journal} {\bibinfo
   {journal} {Phys. Rev. D}\ }\textbf {\bibinfo {volume} {102}},\ \bibinfo
  {pages} {023021} (\bibinfo {year} {2020})},\ \Eprint
  {https://arxiv.org/abs/2004.08293} {arXiv:2004.08293 [astro-ph.HE]}
  \BibitemShut {NoStop}%
\end{thebibliography}%
\end{document}